\documentclass[prb,amsmath,amssymb,twocolumn]{revtex4-1}
\usepackage[pdftex]{graphicx}
\usepackage[pdftex]{epsfig}
\usepackage{epstopdf}
\usepackage{dcolumn}
\usepackage{bm}
\usepackage{mciteplus}
\usepackage{color}

\newcommand{\units}[1]{\ensuremath{\mathrm{#1}}}
\newcommand{\amount}[2]{\ensuremath{#1\:\,\units{#2}}}
\newcommand{\VSD}[0]{\ensuremath{V_{\mathrm{SD}}}}
\newcommand{\VM}[0]{\ensuremath{V_{\mathrm{M}}}}
\newcommand{\VL}[0]{\ensuremath{V_{\mathrm{L}}}}

\newcommand{\V}[1]{\ensuremath{V_{\mathrm{#1}}}}

\begin{document}

\title{Single-shot measurement and tunnel-rate spectroscopy of a
  Si/SiGe few-electron quantum dot}

\author{Madhu Thalakulam$^1$}
\email{mthalak@sandia.gov}
\author{C. B. Simmons$^2$}
\author{B. J. Van Bael$^2$}
\author{B. M. Rosemeyer$^2$}
\author{D. E. Savage$^2$}
\author{M. G. Lagally$^2$}
\author{Mark Friesen$^2$}
\author{S. N. Coppersmith$^2$}
\author{M. A. Eriksson$^2$}
\affiliation{$^1$MS 1415, Sandia National Laboratories, Albuquerque,
New Mexico 87185, USA}
\affiliation{$^2$University of Wisconsin-Madison, Madison, Wisconsin 53706, USA}

\begin{abstract}
We investigate the tunnel rates and energies of excited states of small numbers of electrons in a quantum dot fabricated in a Si/SiGe heterostructure.  Tunnel rates for loading and unloading electrons are found to be strongly energy dependent, and they vary significantly between different excited states.  We show that this phenomenon enables charge sensing measurements of the average electron occupation that are analogous to Coulomb diamonds.  Excited-state energies can be read directly from the plot, and we develop a rate model that enables a quantitative understanding of the 
{relative sizes of different}
electron tunnel rates.
\end{abstract}

\maketitle

\newpage

\section{Introduction}
Many readout schemes for solid-state quantum computing
architectures rely on accurate control and
real-time measurement of electron tunneling.\cite{Loss:1998p120,Kane:1998p133,Elzerman:2004p431,Petta:2005p2180,PioroLadriere:2008p776}  In GaAs quantum dots, tunneling rates have been tuned by controlling gate voltages as well as by exploiting energy-dependent tunneling,\cite{MacLean:2007p1499}  enabling the identification of both orbital and Zeeman excited states.\cite{Elzerman:2004p731,Schleser:2005p035312,Amasha:2008p1500}
Single-shot readout using spin-to-charge conversion has been performed by using a capacitively coupled quantum point contact (QPC) as the charge sensor.\cite{Elzerman:2004p431,Hanson:2005p719,Amasha:2008p1987,Barthel:2009p160503}

In silicon, the effects of energy dependent tunneling are expected to
be enhanced compared to GaAs, because of the carriers' larger effective mass.
Time-averaged measurements of spin-dependent electron tunneling into quantum dots have been made using quantum point contacts.\cite{Xiao:2010p1876,Hayes:2009preprint}  Single electron transistors have been used to measure tunnel rates\cite{Huebl:2010p1868} and to perform single-shot spin readout of electrons on individual dopants.\cite{Morello:2010p2645}

Here we investigate the tunnel rates and energies of excited states of small numbers of electrons in a quantum dot fabricated in a Si/SiGe heterostructure.  We find tunnel rates for loading and unloading electrons are strongly energy dependent: states below the Fermi level load more slowly as their
energy decreases, and states above the Fermi level unload more rapidly
as their energy increases.  Single-shot measurement of these tunnel events is achieved with 
{good charge sensitivity.}
We further show that loading and unloading
tunnel rates vary significantly between different excited states.
This phenomenon enables charge sensing measurements to produce plots
analogous to Coulomb diamonds.  The use of charge sensing to create such plots enables calibration of the gate voltage-to-energy ratio $\alpha$ without the need for measurable transport through the quantum dot itself. Excited-state energies can be read
directly from the plot, and we develop a rate model to extract quantitative 
{relations between}
the tunnel rates from the experimental measurements. A
simulated map of charge sensing measurements as a function of
source-drain and gate voltages agrees well with the experimental
data. We find significant variations in tunnel rates that may be
useful in the loading and measurement of spins in quantum dots.

\section{Methods}
The sample used in this work was grown by chemical vapor deposition on
a Si(001) substrate with phosphorus doping of
1-10~$\Omega\cdot\text{cm}$, which was polished 2$^\circ$ towards
[010].\cite{Thalakulam:2010p183104}  The structure was step-graded to
the composition Si$_{0.64}$Ge$_{0.36}$.  The Ge concentration was then
reduced to Si$_{0.68}$Ge$_{0.32}$ and a 1~$\mu$m buffer layer was
grown, resulting in a final strain relaxation of 95\%. On top of the
relaxed SiGe, we grew 18~nm of Si (the quantum well), 22~nm of undoped
Si$_{0.68}$Ge$_{0.32}$, 1~nm of doped Si$_{0.68}$Ge$_{0.32}$, 45~nm of
undoped Si$_{0.68}$Ge$_{0.32}$, and a 9~nm Si cap layer.  The top
gates were formed by electron beam evaporation of Pd onto the
HF-etched surface of the heterostructure.\cite{Slinker:2005p246,Berer:2006p162112}  The gates sit on a square
mesa of width 35~$\mu$m that was defined by reactive ion etching.  For
the experiment reported here, the sample was illuminated with red
light for 10~s at a temperature of 4.2~K, before cooling to a
refrigerator base temperature of 20~mK.  Magnetoresistance
measurements obtained from the same heterostructure give a carrier
density of $5.15\times10^{11}$~cm$^{-2}$ and a mobility of
$120,000$~cm$^{2}$/Vs (after illumination).  Data are reported in this
paper with the quantum dot in both the
one-electron\cite{Thalakulam:2010p183104,Ciorga:2000p16315,Simmons:2007p213103}
and many electron regimes.

\section{Results}
\begin{figure}[t]
\includegraphics[width=8cm]{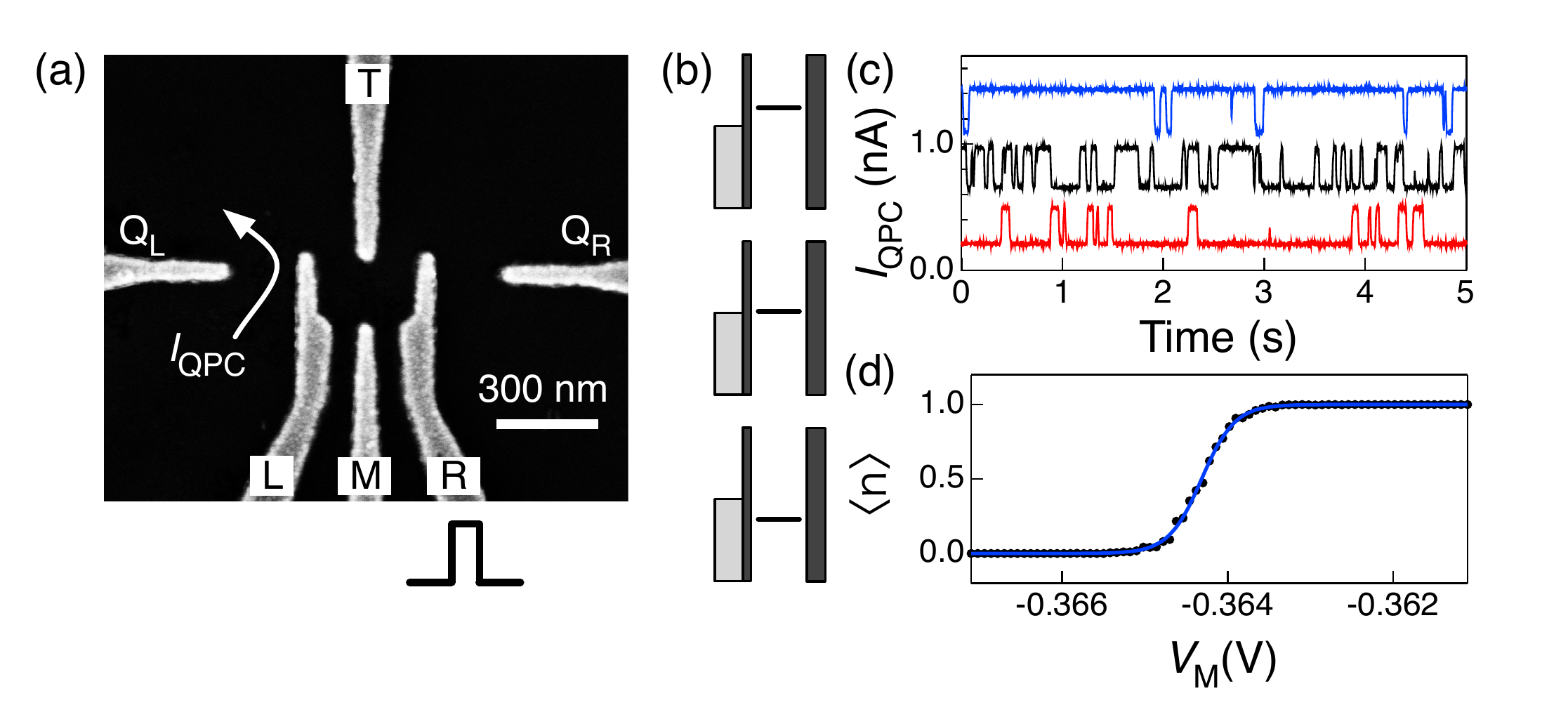}
\caption{\label{fig:sample} (Color online) (a) Scanning electron micrograph of
    the quantum dot top-gates, with gate labels shown.  For
    pulsed-gate measurements, a pulse sequence was applied to gate R.
    (b) Chemical potential diagrams for tunneling between the left
    reservoir and the quantum dot.  Loading (top diagram) and
    unloading (bottom diagram) are achieved by varying the gate
    voltages \VM\ or \V{R}.  (c) Typical oscilloscope traces of
    $I_{\mathrm{QPC}}$ versus time, showing real-time tunneling
    events.  Downward edges indicate loading, upward edges indicate
    unloading.  The three traces correspond to the three energy
    configurations shown in (b).  (d) Fractional dot occupation as a
    function of \VM, obtained from traces like those in (c).}
\end{figure}

\begin{figure}[t]
\includegraphics[width=8cm]{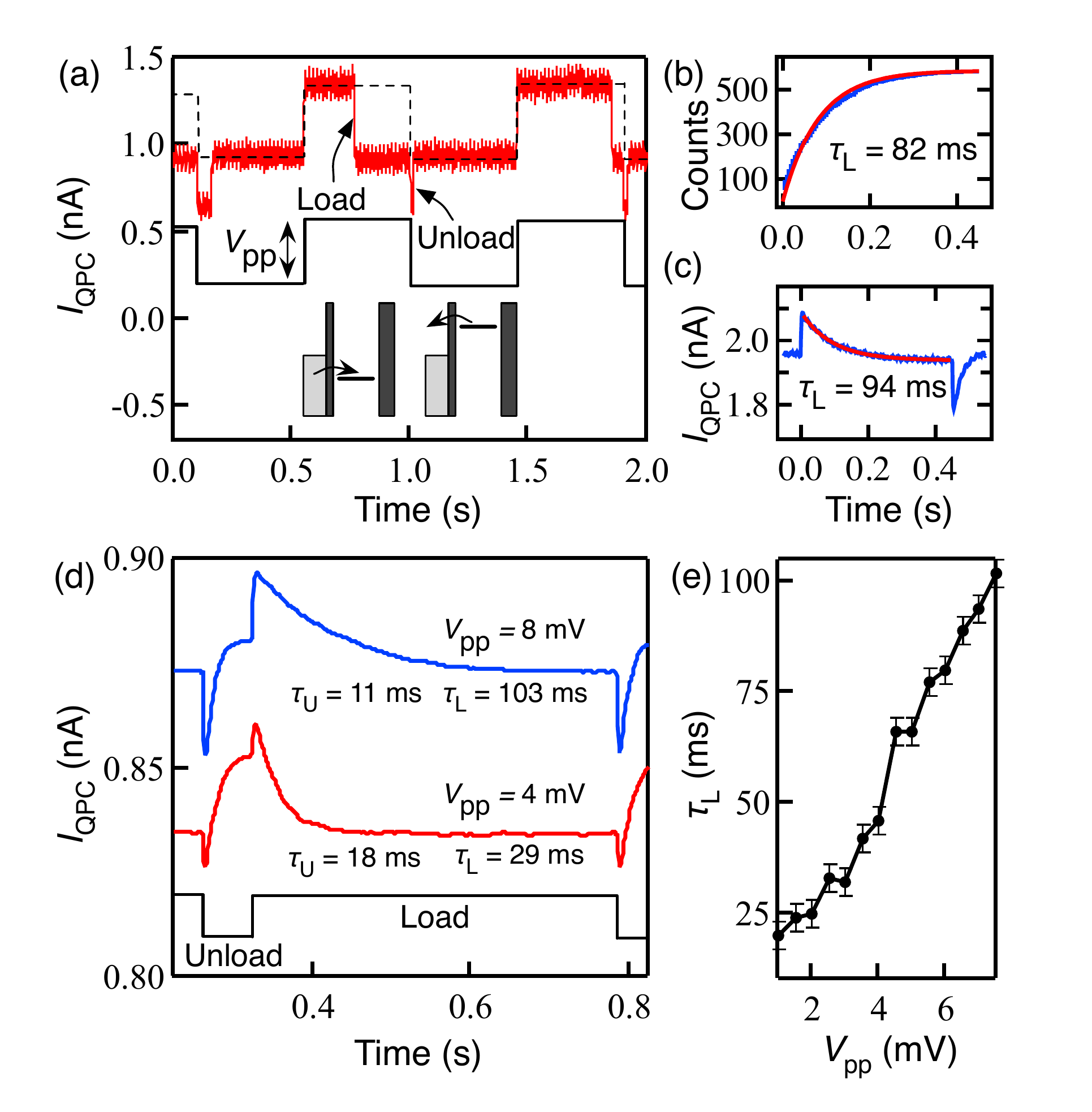}
\caption{\label{fig:pulses} (Color) Pulsed gate voltage measurements.
    (a) A typical measurement of $I_{\mathrm{QPC}}$ (red curve),
    together with its contribution from cross talk (dashed black
    curve). The voltage pulse sequence applied to gate R is shown as
    the solid black curve (arb.\ units).  Insets: chemical potential
    diagrams for loading and unloading.  (b) 
   {Cumulative counting statistics} 
    for loading events vs.\ time with
    \amount{V_{\mathrm{pp}}=8}{mV}. (c) Average of 600 current traces
    (blue curve) and an exponential fit (red curve).  (d) Average of
    512 current traces (red \& blue curves), for two different values
    of $V_\mathrm{pp}$ (the corresponding pulse sequence is shown in
    black with arb.\ units). 
    (e) Loading time vs.\ pulse
    amplitude, showing a strong increase in tunneling time as the
    quantum dot chemical potential is decreased.}
\end{figure}

Fig.~\ref{fig:sample}(a) shows a scanning electron micrograph of the gate
structure of a device identical to the one reported here.  The quantum
dot in this experiment was formed by applying negative voltages to
gates L, M, R, and T.  A charge sensing QPC was formed by applying a
negative voltage to gate
Q$_{\mathrm{L}}$.\cite{Sakr:2005p223104,Simmons:2007p213103,Nordberg:2009p202102}
The QPC was biased at 100~$\mu$V, and the drain current was monitored
with a low-noise current preamplifier. For the data shown in
Figs.~\ref{fig:sample} and \ref{fig:pulses}, the quantum dot barriers and
gate voltages were tuned such that tunneling occurred predominately
through the left barrier, and the dot occupation could be tuned
between zero and one.  $I_{\mathrm{QPC}}$ was high when the dot was
empty and low when the dot was occupied.

As shown in Fig.~\ref{fig:sample}(c), monitoring the QPC current as a
function of time enables measurement of single-electron tunneling on
and off the dot.  In the absence of a source-drain voltage \VSD\
across the dot, the charge occupation is determined solely by the
alignment of the electron chemical potential in the quantum dot, which is controlled by
the voltage \VM\ on gate M, with the Fermi level of the leads, as
depicted in Fig.~\ref{fig:sample}(b). Three typical time traces of
$I_\mathrm{QPC}$ are shown in Fig.~\ref{fig:sample}(c), corresponding to
the three energy configurations in panel (b).  In the top trace of
panel (c), the chemical potential of the dot is slightly above the
Fermi level, and the dot is almost always unloaded, except for
occasional thermal fluctuations.  The bottom trace corresponds to the
opposite situation, and the middle trace corresponds to the case where
the chemical potentials of dot and lead are nearly aligned, leading to
an average dot occupation of 50\%.  
{Based on a signal-to-noise ratio of 32 and a preamplifier rise time of \amount{3}{ms}, we estimate a charge sensitivity of \amount{2.9\times 10^{-3}}{e/\sqrt{Hz}} for our QPC measurements.}

As shown in Fig.~\ref{fig:sample}(d), data of this type enables a
measurement of the Fermi-Dirac distribution for the dot occupation
$\langle n(\VM) \rangle = f_D(\mu_g)$, where $\mu_g$ is the chemical
potential of the dot ground state.  By repeating measurements like
that shown in panel (d) for a series of increasing temperatures, we
acquire a set of data that can be used to determine the
proportionality constant $\alpha$ between the chemical potential and
\VM\ for the one-electron dot.\cite{Simmons:2009p3234} From a global
fit we determine $\alpha=0.129\pm 0.004$~{meV/mV}, where the error bar
is determined by the quality of the fit.  
{The electron temperature is estimated as \amount{298}{mK}.}

By applying pulsed gate voltages to gate R, we can repeatably load and
unload the quantum dot, enabling measurement of the loading and
unloading tunnel rates as a function of the quantum dot chemical
potential.  A typical pulse sequence is shown in Fig.~\ref{fig:pulses}(a)
(solid black trace).  Loading occurs during the high portion of the
gate voltage cycle, while unloading occurs during the low portion, as
indicated in the insets.  
{The tunnel barriers can be controlled independently of the dot chemical potential, although cross talk is always present.  Here, we tune the tunnel barriers by adjusting the voltage, primarily on gate L,}
so that the tunneling times are much longer than the rise
time of the current preamplifier in the QPC circuit.  
{After this initial adjustment, gate L is held fixed, while the voltage is varied on gate R.}

A typical
measurement trace corresponding to loading and unloading the dot is
shown in red in Fig.~\ref{fig:pulses}(a).  Steps in the traces occur in
pairs; the first step edge is precisely correlated with the voltage
pulse and is the cross talk response of the QPC to the voltage pulse
on gate R.  The cross talk component of $I_\mathrm{QPC}$ is sketched
as a dashed black line in Fig.~\ref{fig:pulses}(a).  The second step edge
occurs in between the voltage pulses.  This signal corresponds to a
tunneling event --- either the loading or unloading of the dot.  By
recording the time interval between the voltage pulse and the charging
event, we obtain a direct, single-shot measurement of the loading
time.

Figure~\ref{fig:pulses}(b) shows a 
{plot of the cumulative number of
loading events} 
as a function of the loading time, obtained from a
pulse sequence of identical load-unload cycles with
\amount{V_\mathrm{pp}=8}{mV}.  Loading times are extracted from
$I_\mathrm{QPC}(t)$ in a two step process: first, we check that the
dot is unloaded immediately prior to the loading pulse; second, we
extract the first time that $I_\mathrm{QPC}$ returns to the level
corresponding to a loaded electron.  The results shown in Fig.~\ref{fig:pulses}(b) 
take into account 584 out of 600 such current traces; 16 traces
did not have current levels that could be mapped to the analysis
pattern and were ignored.  As expected, the counts
saturate exponentially, with a rise time of \amount{\tau_L=82 \pm
  9}{ms}. Figure~\ref{fig:pulses}(c) shows a direct average of all 600
current traces.  The decay is well fit by an exponential with time
constant \amount{\tau_L=94 \pm 7}{ms}, comparable to the value
obtained by event counting.

\begin{figure}[t]
\includegraphics[width=8cm]{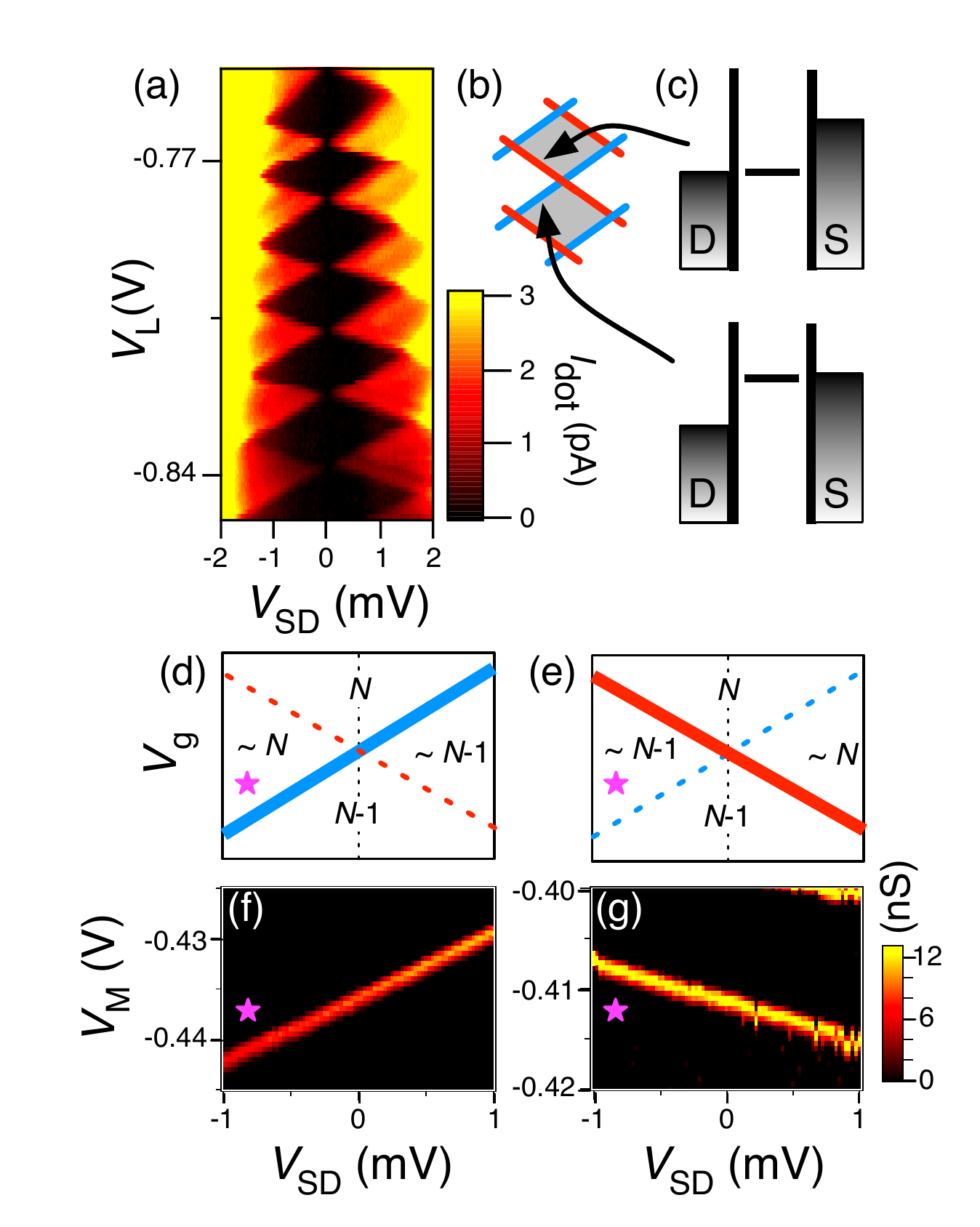}
\caption{\label{fig:diamonds} (Color online) (a) Coulomb diamond plot of
    current $I_{\mathrm{dot}}$ through the quantum dot in the many
    electron regime.  (b) Schematic representation of two Coulomb
    diamonds.  (c) Energy level diagrams showing the dot chemical
    potential and the Fermi level of the leads for two of the edges in
    (b). (d, e) Expected transconductance (solid lines) of the charge
    sensing QPC when the $D$ and $S$ tunnel barriers, respectively,
    are the transport bottleneck.  (f, g) Color scale plot of the
    transconductance $g$ of the QPC obtained as a function of \VM\ and
    \VSD.  Peaks in $g$ represent transitions where dot occupation
    changes. \amount{\V{R}=-0.42}{} and \amount{-0.44}{V} in (f) and
    (g) respectively.
    \VM\ differs in the two panels, to compensate
    for the difference in \V{R}.}
\end{figure}

We measure the loading and unloading rates over a wide range of
$V_{\mathrm{pp}}$.
We use a pulse sequence with an asymmetric duty cycle, shown in black
in Fig.~\ref{fig:pulses}(d), because the unloading times are significantly
faster than the loading times.  The red and blue curves in
Fig.~\ref{fig:pulses}(d) correspond to two different values of the pulse
height $V_\mathrm{pp}$ and are averages of 512 individual
$I_\mathrm{QPC}$ traces.  Loading and unloading times are obtained by
performing exponential fits to the averaged data.
Fig.~\ref{fig:pulses}(e) shows the resulting loading
times as a function of $V_\mathrm{pp}$.  We observe an approximate
five-fold increase in the loading time as $V_\mathrm{pp}$ varies from
$1$ to \amount{6.5}{mV}, corresponding to a shift of about
\amount{343}{\mu eV} in the dot chemical potential during the loading
phase of the cycle.

{The dependence of the tunnel rate on gate voltage is consistent with a simple model of transport in a quantum dot known as energy-dependent 
tunneling.\cite{MacLean:2007p1499,Amasha:2008p1500,Amasha:2008p1987,Simmons:2010p245312}  
In this model, the tunnel rate depends exponentially on the relative height of the tunnel barrier compared to the chemical potential of the dot, with a larger chemical potential corresponding to a higher tunnel rate.  
This explains the trends of the tunnel rate with respect to variations in the gate voltage.  It also explains why the unloading rates are much faster than the loading rates in our dot.
It is important to note that cross talk between the plunger gate and the tunnel barrier would tend to have the opposite effect as energy-dependent tunneling, by raising the tunnel barrier and lowering the tunnel rate.}

{Up to this point, we have studied the tunnel rate between the lead and the ground state of the dot, which varies as a smooth function of the plunger gate voltage, as indicated in Fig.~\ref{fig:pulses}(e).  
We now describe how the tunnel rates may change more abruptly, when excited states enter the bias window.
These excited states may have tunneling matrix elements much larger than the ground state, which depend on the shape and the symmetry of the wavefunctions.\cite{Hanson:2005p719,Ono:2002p1313,Johnson:2005p925,Petta:2005p161301,Johnson:2005p165308} 
In turn, this opens up new doors for measuring the tunnel rates.
Specifically, it allows us to replace gate pulsing with time-averaged current measurements.}
Time-averaging is normally associated with methods where the current flows directly through the quantum dot.
One example is the so-called Coulomb
diamond plot, where the transport current is measured as a function of source-drain bias \VSD~and the plunger gate voltage.\cite{Kouwenhoven:1997p1384} 
Such transport measurements are challenging in the few-electron regime, where
charge sensing is the preferred experimental
technique.\cite{Field:1993p1477}  
We show here that charge sensing techniques enable spectroscopy of silicon quantum dots, with direct correspondence to Coulomb diamond measurements.
Similar measurements were previously obtained in GaAs by Schleser \emph{et
  al}.\cite{Schleser:2005p035312}

Figure~\ref{fig:diamonds}(a) shows a Coulomb diamond plot of the current
through the dot as a function of $V_\text{SD}$ and $V_\text{L}$, for the case
of many-electron occupation and large current flow through the quantum
dot.  Current flows only when the chemical potential of the dot
$\mu_{\mathrm{g}}$, which depends linearly on \VL, lies between the
Fermi energies of the source and drain ($\mu_S$ and $\mu_D$,
respectively).  The black regions in Fig.~\ref{fig:diamonds}(a) are
blockaded.  As $\mu_{\mathrm{g}}$ is lowered, it becomes level with
$\mu_S$; this condition corresponds to the edge of the diamond with
positive slope (blue line in Fig.~\ref{fig:diamonds}(b)), and current
begins to flow.  When $\mu_D\leq \mu_{\mathrm{g}} \leq \mu_S$, current
flows, and the dot occupation alternates between $N$ and $N-1$.  The
condition $\mu_{\mathrm{g}}=\mu_D$ corresponds to the edge of the
diamond with negative slope (red line in Fig.~\ref{fig:diamonds}(b)).  When
$\mu_{\mathrm{g}}<\mu_D$, the electron does not have enough energy to
exit the dot.  In this case, the electron occupation becomes fixed at
value of $N$, and the current is blocked.

Because of the variation in the tunnel rate with gate voltage and the
available energy levels, the same information available in a Coulomb
diamond transport measurement can be derived from time-averaged charge
sensing measurements.  
Here, we study this effect in the several-electron regime.  Figures~\ref{fig:diamonds}(f)
and (g) show plots of the transconductance $g=\partial
I_\text{QPC}/\partial \VM$.  When monitoring the charge sensing QPC
current in this way, the physical picture of Coulomb blockade is
unaffected; however, instead of measuring the current through the
quantum dot, which is immeasurably small in this regime, the QPC
measures the average dot occupation.  In the regime where dot
occupation alternates between $N-1$ and $N$, the average occupation
depends on the electron dwell time, which depends on the loading and
unloading tunnel rates.

The interesting case for charge sensing is when $\VSD \neq 0$ and the
dot chemical potential is in the bias window.  In this regime, the
average fractional occupation $f$, the fractional part of the dot
occupation $\langle n \rangle$, is given by $f
=\Gamma_{gS}/(\Gamma_{gS}+\Gamma_{gD})$, where $\Gamma_{gS}$ is the
tunnel rate from the source to the $N$-electron ground state, and
$\Gamma_{gD}$ is the corresponding rate to the drain. At the location
of the purple stars in Figs.~\ref{fig:diamonds}(d)-(g), if the tunnel barriers
are tuned such that $\Gamma_{gS}\simeq \Gamma_{gD}$, the average
fractional occupation will be $f \simeq 0.5$.  However, it is more
common that the tunnel rates are dissimilar, and when one barrier is
much larger than the other, it becomes a bottleneck for transport.

For the limiting case $\Gamma_{gS}\gg \Gamma_{gD}$, shown in
Figs.~\ref{fig:diamonds}(d) and (f), the electron tunnels into the dot very
quickly when $\mu_{\mathrm{g}}\leq \mu_S$, but it tunnels out very
slowly, so the average occupation will be $\sim N$.  A filling
transition is therefore observed when $\mu_{\mathrm{g}}=\mu_S$, but
not when $\mu_{\mathrm{g}}=\mu_D$.  Thus, the transition occurs along
the edge of the Coulomb diamond with positive slope.  In the opposite
limit, $\Gamma_{gD}\gg \Gamma_{gS}$, shown in Figs.~\ref{fig:diamonds}(e) and
(g), and achieved by changing gate voltage \V{R}, the dot empties so
quickly that the average occupation is $\sim(N-1)$.  In this case, the
filling transition occurs when the dot chemical potential is aligned
with the drain, corresponding to the edge of the Coulomb diamond with
negative slope.  As is clear from Figs.~\ref{fig:diamonds}(f) and (g), the tunnel
barriers are easily tuned into either regime.

\begin{figure}[t]
\includegraphics[width=8.5cm]{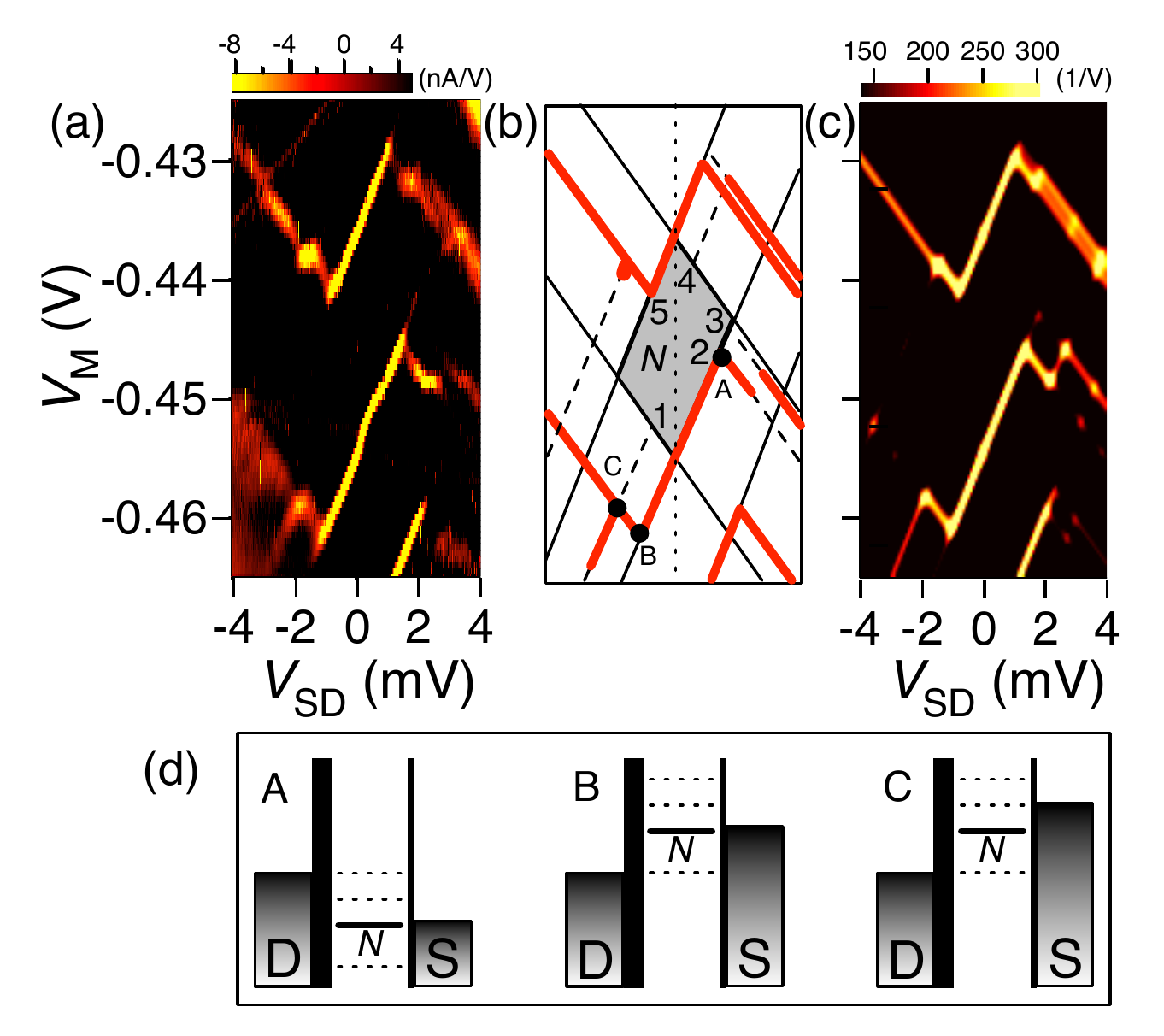}
\caption{\label{fig:zorro} (Color online) Excited-state spectroscopy.  (a)
    Color-scale plot of $g$ as a function of \VSD\ and \VM.  The
    pattern in the transconductance is analogous to a traditional
    Coulomb diamond plot of the differential conductance through a
    quantum dot.  (b) Schematic representation of the ``transition
    map'' observed in (a), showing the role of excited states and how
    the energies of these states can be extracted in analogy with a
    Coulomb diamond plot.  (c) Theoretical reconstruction $\partial f
    /\partial V_{\mathrm{M}}$ of the transition map, based on fitting
    to a rate equation model 
    {described in the Appendix,} 
    demonstrating the robustness of the
    interpretation of the data.  (d) Schematic energy level diagrams
    of the dot at points A, B, and C in panel (b).}
\end{figure}

\begin{table*}[t]
  \caption{Excitation energies and tunnel rates $\Gamma_{ij}$, as extracted from Fig.~\ref{fig:zorro}.  
  Here, $ij$ specifies the tunnel channel, with $i$ corresponding to the quantum dot orbital, 
    and $j$ corresponding to
    the source S or drain D lead. The tunnel rates are obtained 
    by fitting the transition map 
    in Fig.~\ref{fig:zorro}(a) to a finite temperature rate model
    {described in the Appendix.}
    The excitation energies (1-5) are measured relative to the ground state ($g$).
    This procedure provides tunnel rates that are normalized to the $gS$ tunnel channel.
   {The tunnel rates are then calibrated with respect to the $gD$ tunnel channel, as described in the text.}}
\centering
\begin{tabular}{c c c c c c}
  \\ \hline\hline
  Filling & Dot & Energy & Tunnel & Normalized tunnel & Calibrated tunnel \\
  transition & state & (meV) & channel & rate, $\Gamma_{ij}/\Gamma_{gS}$ & rate, $\Gamma_{ij}$ ($10^6$~s$^{-1}$) \\ [0.3ex]
  \hline
  $N \leftrightarrow (N-1)$ & $g$ & 0 & $S$ & 1 & 4.7\\
  & $g$ & 0 & $D$ & 0.15 & 0.7 \\
  & 1 & 0.92 & $S$ & 6 & 28 \\
  & 2 & 1.43 & $D$ & 2.3 & 11 \\
  & 3 & 1.90 &  $D$ & 10 & 47 \\
  \hline
  $N \leftrightarrow (N+1)$ & $g$ & 0 & $S$ & 1 & 3.5 \\
  & $g$ & 0 & $D$ & 0.2 & 0.7\\
  & 4 & 0.69 & $S$ & 1.2 & 4.2\\
  & 5 & 0.82 & $D$ & 6 & 21 \\
  \hline\hline
\end{tabular}
\label{table1}
\end{table*}

Positive and negative slopes can also be observed without retuning the
tunnel barriers, simply by expanding the bias window, as shown in
Fig.~\ref{fig:zorro}(a).  Here, we have chosen the tuning $\Gamma_{gS} >
\Gamma_{gD}$.  The mapping between the filling transitions and the
conventional Coulomb diamonds is sketched in Fig.~\ref{fig:zorro}(b), where
one of the diamonds has been shaded gray.  We observe sharp corners in
the transition map, which we attribute to excited states entering the
bias window, similar to lines corresponding to excited states in a
Coulomb diamond plot.  Three such transitions are labelled A, B, and C
in panel (b).  Near $\VSD=0$, the charge transition line has a
positive slope, since the $D$ barrier forms the bottleneck, and the
filling transition occurs when the dot level is aligned with the
source.  In this range of \VSD, all excited states lie outside the
bias window, and the average fractional occupation
$f=\Gamma_{gD}/(\Gamma_{gD}+\Gamma_{gS})\simeq 0$.  
{(See Appendix for details.)}
When an excited
state $x$ enters the window, such as at point A in Fig.~\ref{fig:zorro}(b),
the occupation is given by
$f=(\Gamma_{gD}+\Gamma_{xD})/(\Gamma_{gD}+\Gamma_{gS}+\Gamma_{xD})\simeq
1$.  In this case, the approximate equality holds because of the
strong tunnel coupling between the drain and the excited state: 
$\Gamma_{xD}\gg \Gamma_{gS} \gg \Gamma_{gD}$.  Thus, a switch in occupation is caused
by a switch in the rate-limiting tunnel barriers.  We can understand the
changes in slope near points~B and C by similar arguments.  The
transition map expands in piece-wise fashion, with sharp changes in
slope indicating the presence of excited states that change which
barrier corresponds to the bottleneck tunnel rate.  Such changes in
slope are visible over
a wide range in \VSD\ and \VM.  Excited states with tunnel couplings
that do not alter the time averaged dot occupation will be invisible.

We can analyze the experimental data of Fig.~\ref{fig:zorro}(a)
using a finite temperature rate model, 
{as described in detail in Appendix~A.}
In brief, the rate equations are
obtained by first expressing the currents for individual transport
channels.  For example, the tunnel current
from the source to the ground state of the dot is given by $I_{gS}=-e\Gamma_{gS}
f_S(\mu_{\mathrm{g}})(1-f)$, where $f_S(\mu_{\mathrm{g}})$ is the
Fermi function for the source reservoir, evaluated at the chemical
potential of the dot ground state.  By enforcing current conservation
and assuming an infinite decay rate for excited states, we obtain an
equation for $f$ as a function of tunnel rates, excited-state
energies, and temperature.  It is important to note that, although an
electron from the drain can tunnel into an $N$-electron excited state,
it cannot tunnel out through the same channel.  Instead, the electron
decays very quickly and tunnels out through the ground
state.\cite{Hanson:2007p720}  The decay process causes an intrinsic
asymmetry, one which favors loading rather than unloading.

The quantity $\partial f/\partial \VM$ corresponds directly to the
transconductance of Fig.~\ref{fig:zorro}(a).  We fit this result to the
experimental data, obtaining the parameters shown in the next-to-last column of Table~I, as well as the
theoretical reconstruction of the transition map shown in
Fig.~\ref{fig:zorro}(c).  In the fitting procedure, no correlations were
assumed between the excited states in the different transport
channels. We estimate an uncertainty of about \amount{60}{\mu eV} for
the excited-state energies, and an uncertainty of about 20\% for these relative
tunnel rates.  

{
Since transport currents are immeasurably small in this experimental configuration,
we are not able to determine the tunnel rates directly.
In the next-to-last column of Table~I, the tunnel rates are normalized relative to the $gS$ tunnel channel, as described in the Appendix.
Here, the tunnel rates for the $N\leftrightarrow(N-1)$ and $N\leftrightarrow(N+1)$ processes are normalized independently, since they involve different gate voltages, and should not be exactly equal.}

{
We were able to obtain an approximate calibration for the tunnel rates, however, by retuning the quantum dot to allow pulsing experiments.
Shortly after the data in Fig.~\ref{fig:zorro} was obtained, the right tunnel barrier (R) was pinched off, allowing electrons to tunnel through just the left ($D$) barrier.
The tunnel barrier gate voltage $V_L$ was left essentially unchanged.
Tunnel rates corresponding to the $gD$ process were then acquired on two successive days, with variations on the order of 25\%.
In this way we obtain the estimates $\Gamma_{gD}=5.4\times 10^5$~s$^{-1}$ and $8.5\times 10^5$~s$^{-1}$ for the $N\leftrightarrow(N-1)$ loading and unloading processes, respectively.
The difference between loading and unloading rates is comparable to those described above, and it is also consistent with predictions of energy dependent tunneling.
In the final column of Table~\ref{table1}, we use the mean tunnel rate estimate $7.0\times 10^5$~s$^{-1}$ to calibrate the relative rates listed in the next-to-last column.
The same calibration provides a rough estimate for the $N\leftrightarrow(N+1)$ tunneling process, and we provide the corresponding results in Table~\ref{table1}.}

The lever arm $\alpha$, which converts gate voltage to dot energy, is
usually extracted from a Coulomb diamond plot.  
Here, we can extract $\alpha
= 0.095 \pm 0.004$~{meV/mV} from Fig.~\ref{fig:zorro}(a), even though transport
through the quantum dot itself is immeasurably slow.  The uncertainty
in $\alpha$ is determined by the resolution of the data plot in both
\VSD\ and \VM.  This value of $\alpha$ is different than that quoted
above when the quantum dot was occupied by a single electron, because
the dot has been retuned and now is occupied by several electrons.

\section{Summary}
In summary, tunnel rates in and out of a few-electron Si
quantum dot were measured by
single-shot charge sensing.
{The rates were shown to depend strongly on gate voltage, in a manner opposite to that expected from cross talk, and consistent with energy-dependent tunneling.}  
We also have shown
that a map of the time averaged dot occupation, obtained by charge
sensing, provides direct spectroscopic information about the quantum
dot.  Further, energy calibration --- the determination of $\alpha$
--- does not require retuning of the tunnel barriers, because both the
positive and negative slopes, which are usually extracted from Coulomb
diamonds, are visible in a plot like that shown in Fig.~\ref{fig:zorro}(a).
The sharp corners observed in the transition map are attributed to
excited states entering the bias window and their effect on the
tunneling bottleneck.  Sudden switching of the bottleneck is expected,
because the tunneling matrix elements for excited states can be large,
and because there is a built-in asymmetry in the tunneling process,
due to the fast decay of excited states.

\section*{Acknowledgements}
We acknowledge support from ARO and LPS (Grant No.\ W911NF-08-1-0482),
NSF (Grant No.\ DMR-0805045), United States Department of Defense, and
DOE (Grant No.\ DE-FG02-03ER46028). The US government requires publication of the following disclaimer:  the views and conclusions
contained in this document are those of the authors and should not be
interpreted as representing the official policies, either expressly or
implied, of the US Government.  This research utilized NSF-supported
shared facilities at the University of Wisconsin-Madison.  Sandia
National Laboratories is a multi-program laboratory managed and
operated by Sandia Corporation, a wholly owned subsidiary of Lockheed
Martin Corporation, for the U.S. Department of Energy's National
Nuclear Security Administration under contract DE-AC04-94AL85000.

\appendix
\section{Rate equations for the transition map}
In this Appendix we present a more detailed explanation of the Coulomb diamond-like behavior observed in the transconductance data of Fig.~\ref{fig:zorro}(a).
As explained in the main text, the presence of switches in the transition map indicates that the source ($S$) and drain ($D$) tunnel barriers are asymmetric.  A change of slope occurs whenever the slow barrier (the bottleneck) switches from $S$ to $D$, or visa versa, due to excited states entering the bias window.\cite{Schleser:2005p035312}
Here, we will derive a simple theoretical description of the switching observed in this transition map.

\begin{figure}[t] 
  \centering
  \includegraphics[width=3in,keepaspectratio]{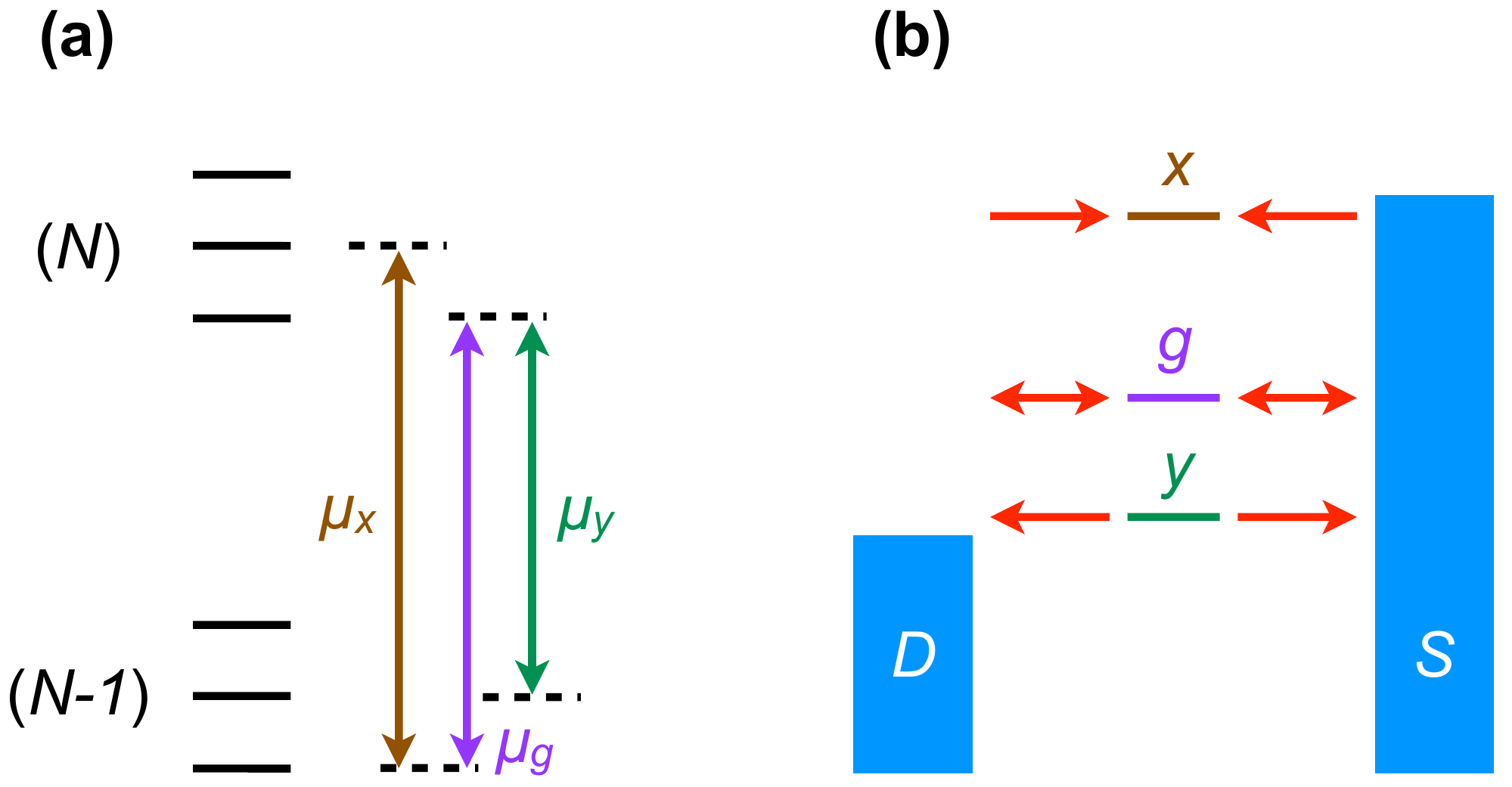}
  \caption{(Color online)
  (a) Ground ($g$) and excited ($x,y$) state transitions in a quantum dot, between the electron occupations $(N-1)$ and $N$.  The chemical potential describing transitions between the ground states is $\mu_g$.  Transitions to excited states may occur in either direction:  when the dot loads ($\mu_x$) or unloads an electron ($\mu_y$).
  (b) The types of tunneling processes considered in this work.  Fast decay of excited states essentially prohibits $x$-type unloading or $y$-type loading processes.}
  \label{fig:combo}
\end{figure}

The transition between $(N-1)$ and $N$-electron states in a quantum dot may involve excited orbitals in either the $(N-1)$ or $N$-electron manifolds.  
The chemical potentials for several such processes are sketched in Fig.~\ref{fig:combo}(a).
There are two necessary ingredients for observing a single change of slope.  
(1) One of the barriers (say, $D$) must form a bottleneck for transitions between the $(N-1)$ and $N$-electron ground states; in other words, $\Gamma_{gD} \ll \Gamma_{gS}$, where $\Gamma$ signifies a tunnel rate.  
(2) There must be an excited state (not necessarily the lowest excited state) for which the tunnel rate to $D$ is faster than the ground state tunnel rate to $S$; in other words, $\Gamma_{xD} \gg \Gamma_{gS}$.  
The fact that the decay rate $\Gamma_{xg}$ between the excited and ground states of the quantum dot is much faster than any of the tunnel rates also facilitates switching.
Additional changes of slope may occur when new excited states enter the bias window.
Note that energy-dependent tunneling is not a leading order effect in the switching behavior 
observed in Fig.~\ref{fig:zorro}(a) and it will not be explicitly considered here.


We will consider each of the different, sequential transition processes shown in Fig.~\ref{fig:combo}(b).  Although the case $\mu_S >\mu_D$ is presented here, analogous processes are also present when $\mu_S <\mu_D$.
We will adopt the notation $x$ to refer processes between the $(N-1)$-electron ground state and an $N$-electron excited state, as sketched in Fig.~\ref{fig:combo}(a).  
Only one such transition is shown in the figure, although there may be many, in practice.  
The notation $y$ refers to processes between the $N$-electron ground state and an $(N-1)$-electron excited state.  
The transition between ground states, $g$, can play a role in both loading and unloading of the quantum dot.
However, $x$ or $y$-type processes are essentially uni-directional, as indicated in panel (b), due to the fact that $\Gamma_{xg}$ is very large.  
Thus, when an $x$-type process occurs, the loaded state immediately decays to the $N$-electron ground state, before the dot has a chance to unload.  
Similarly, a $y$-type process may unload the dot into an $(N-1)$-electron excited state, which then decays to the ground state before loading can occur.  
Transitions between $N$ and $(N-1)$-electron excited states are not forbidden.
However, they are strongly suppressed by the same fast decay process.

\begin{figure*}[t] 
  \centering
  \includegraphics[width=4in,keepaspectratio]{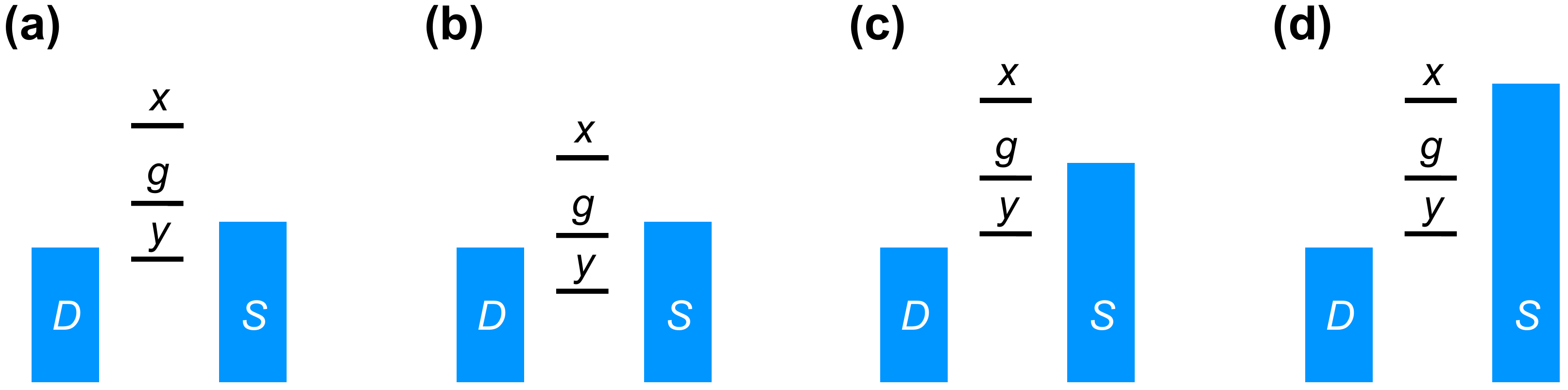}
  \caption{(Color online)
  Cases considered for discussing the quantum dot filling, Eq.~(\ref{eq:big}).}
  \label{fig:cases}
\end{figure*}

We now derive an equation for the steady-state fractional filling of the quantum dot, $f$.  
For simplicity, we will assume that the filling transitions observed in Fig.~\ref{fig:zorro}(a) are thermally broadened, although lifetime broadening may also play a role.\cite{datta}
The elastic loading of the quantum dot ground state from the source is then described by the tunneling current $I_{Sg}$, given by
\begin{equation}
I_{Sg}=e\Gamma_{Sg}f_S(\mu_g)(1-f) .
\end{equation}
Here, $\Gamma_{Sg}$ is the tunnel rate from the source to the $N$-electron ground state, and $f_S(\mu_g)$ is the Fermi function for the source, evaluated at the chemical potential for process $g$.  The reverse process is written as
\begin{equation}
I_{gS}=-e\Gamma_{gS}[1-f_S(\mu_g)]f .
\end{equation}
From here on, we will assume that $\Gamma_{Sg}=\Gamma_{gS}$.

We can write down the currents for all such processes, keeping in mind that a process like $S\rightarrow x$ is allowed, while the reverse is not, due to the fast relaxation of the excited states.
We can obtain a steady-state solution by summing up the currents through the $S$ and $D$ barriers and equating the results.  Solving for $f$, we obtain
\begin{widetext}
\begin{equation}
f = \frac{\Gamma_{gD}f_D(\mu_g) + \Gamma_{gS}f_S(\mu_g) 
+\sum_i \left[ \Gamma_{x_iD}f_D(\mu_{x_i}) + \Gamma_{x_iS}f_S(\mu_{x_i}) \right]}
{\Gamma_{gD} + \Gamma_{gS} 
+\sum_i \left[ \Gamma_{x_iD}f_D(\mu_{x_i}) + \Gamma_{x_iS}f_S(\mu_{x_i}) \right]
+\sum_i \left[ \Gamma_{y_iD}\left( 1-f_D(\mu_{y_i})\right) 
+ \Gamma_{y_iS}\left(1-f_S(\mu_{y_i})\right) \right]} . \label{eq:big}
\end{equation} 
\end{widetext}
Here, we have included all possible excited states, although some of the processes may be invisible, due to weak coupling to the leads.

As an example, we consider the different cases shown in Fig.~\ref{fig:cases}.  
We have focused on the bias $\mu_S>\mu_D$; however, analogous results are obtained for the opposite bias.  
We will consider $\Gamma_{gD}\ll \Gamma_{gS}$, so that the drain barrier forms the bottleneck.  
To simplify the discussion, we will set $T=0$, although the final fitting, shown in Fig.~\ref{fig:zorro}(c), includes temperature as a fitting parameter.

In Figs.~\ref{fig:cases}(a) and (b), we assume that $\mu_S$ is just slightly larger than $\mu_D$.  In particular, $(\mu_S-\mu_D) < (\mu_g-\mu_y)$, so the $g$ and $y$ processes may not be in the bias window simultaneously.  
At $T=0$, the arrangement shown in 
Fig.~\ref{fig:cases}(a) involves no loading processes, as consistent with Fig.~\ref{fig:combo}(b) and the discussion in the main text.
The Fermi functions take the values 0 or 1, and Eq.~(\ref{eq:big}) reduces to $f=0$, indicating an empty dot.
On the other hand, for the arrangement shown in Fig.~\ref{fig:cases}(b), the $g$ process can load and unload. 
In this case, Eq.~(\ref{eq:big}) reduces to 
\begin{equation}
f=\frac{\Gamma_{gS}}{\Gamma_{gS}+\Gamma_{gD}}\simeq 1. \label{eq:flow}
\end{equation}
In this regime, the dot is filled because of the bottleneck at the drain barrier.  
Thus, the filling transition occurs when $\mu_g=\mu_S$.

In Fig.~\ref{fig:cases}(c), the bias is increased such that $(\mu_S-\mu_D) > (\mu_g-\mu_y)$.  As before, when $\mu_g>\mu_S$, Eq.~(\ref{eq:big}) reduces to $f=0$.  
When $\mu_g<\mu_S$, the quantum dot can load through channel $g$, but it can unload through channels $g$ or $y$, as consistent with Fig.~\ref{fig:combo}(b).  
Now Eq.~(\ref{eq:big}) reduces to 
\begin{equation}
f=\frac{\Gamma_{gS}}{\Gamma_{gS}+\Gamma_{gD}+\Gamma_{yD}}\simeq 0,
\end{equation}
where the final equality holds when the tunnel rate for the process $y\rightarrow D$ is large; that is, when $\Gamma_{yD}\gg \Gamma_{gS}$.  When $\mu_y<\mu_D$, we recover Eq.~(\ref{eq:flow}).  
Thus, for the arrangement shown in Fig.~\ref{fig:cases}(c), the filling transition occurs when $\mu_y=\mu_D$.  
The alignment of the filling transition with the drain rather than the source causes a change of slope in the transition map, as discussed in the main text. 
The switching of the slope occurs precisely when $(\mu_S-\mu_D) = (\mu_g-\mu_y)$, corresponding to the excitation energy of the $(N-1)$-electron dot.

In Fig.~\ref{fig:cases}(d), the bias is increased such that $(\mu_S-\mu_D) > (\mu_x-\mu_y)$.  When all three chemical potentials lie inside the bias window, Eq.~(\ref{eq:big}) reduces to
\begin{equation}
f=\frac{\Gamma_{gS}+\Gamma_{xS}}{\Gamma_{gS}+\Gamma_{gD}+\Gamma_{yD}+\Gamma_{xS}}\simeq 1.
\end{equation}
The final equality holds if we assume that $\Gamma_{xS}\gg \Gamma_{yD}$.  
Thus, the filling transition occurs when the $x$ process aligns with the source, again causing a change of slope in the transition map.  
In this case, the change of slope occurs precisely when $(\mu_S-\mu_D) = (\mu_x-\mu_y)$, corresponding to the point where the bias is equal to the \emph{sum} of excitation energies for the $(N-1)$ and $N$-electron dots.

The transition map therefore gives a direct method for performing spectroscopy in a quantum dot, since the changes of slope correspond to excited state transitions entering the bias window.  The corresponding energies can be read off directly from the $V_\text{SD}$ axis in Fig.~\ref{fig:zorro}.  However, the excited states can only be observed if they couple strongly to the lead.  

It is not surprising that consecutive transitions should have larger tunnel rates, since excited states tend to be more spatially extended.  However, it is also possible for excited states to be poorly coupled to the leads.  
Such states will not cause switching, and will be rendered invisible in the transition map.
Similarly, a given level may couple differently to the source and drain reservoirs.  
For one reservoir, it may form an important tunnel channel, while for the other, it may form a bottleneck.

The fitting result shown in Fig.~\ref{fig:zorro}(c) was obtained by simultaneously fitting the tunneling and energy parameters, appearing in the derivative $\partial f/\partial V_M$ of Eq.~(\ref{eq:big}), to the transconductance data of panel (a).  
We have included a ``noise floor," below which all the data were assumed to be indistinguishable from zero, as was done in the experimental plot.  
We also assumed a saturation value for $\partial f/\partial V_M$, corresponding to the color yellow.
Finally, we included temperature as a fitting parameter.

Table~I shows the results of our fitting analysis for one of the Coulomb diamonds, corresponding to the shaded diamond in Fig.~\ref{fig:zorro}(b).
We also obtain the temperature estimate of $T\simeq 0.6$~K.  
This result does not match the estimate obtained from Fig.~\ref{fig:sample}(d), due to differences between conventional and pulsing transport techniques, and the errors introduced by the low and high transconductance cut-offs used in our fitting.

Note that the bright spots in the transition lines with negative slope, at the top right of the transition map in Fig.~\ref{fig:zorro}(c), are real.  They occur along extensions of lines associated with excited states from lower diamonds, as shown in panel~(b).  The corresponding bright spots are offset in panel~(a), due to a charging event which shifted the transition line slightly to the right.

%


\begin{thebibliography}{27}%
\makeatletter
\providecommand \@ifxundefined [1]{%
 \@ifx{#1\undefined}
}%
\providecommand \@ifnum [1]{%
 \ifnum #1\expandafter \@firstoftwo
 \else \expandafter \@secondoftwo
 \fi
}%
\providecommand \@ifx [1]{%
 \ifx #1\expandafter \@firstoftwo
 \else \expandafter \@secondoftwo
 \fi
}%
\providecommand \natexlab [1]{#1}%
\providecommand \enquote  [1]{``#1''}%
\providecommand \bibnamefont  [1]{#1}%
\providecommand \bibfnamefont [1]{#1}%
\providecommand \citenamefont [1]{#1}%
\providecommand \href@noop [0]{\@secondoftwo}%
\providecommand \href [0]{\begingroup \@sanitize@url \@href}%
\providecommand \@href[1]{\@@startlink{#1}\@@href}%
\providecommand \@@href[1]{\endgroup#1\@@endlink}%
\providecommand \@sanitize@url [0]{\catcode `\\12\catcode `\$12\catcode
  `\&12\catcode `\#12\catcode `\^12\catcode `\_12\catcode `\%12\relax}%
\providecommand \@@startlink[1]{}%
\providecommand \@@endlink[0]{}%
\providecommand \url  [0]{\begingroup\@sanitize@url \@url }%
\providecommand \@url [1]{\endgroup\@href {#1}{\urlprefix }}%
\providecommand \urlprefix  [0]{URL }%
\providecommand \Eprint [0]{\href }%
\providecommand \doibase [0]{http://dx.doi.org/}%
\providecommand \selectlanguage [0]{\@gobble}%
\providecommand \bibinfo  [0]{\@secondoftwo}%
\providecommand \bibfield  [0]{\@secondoftwo}%
\providecommand \translation [1]{[#1]}%
\providecommand \BibitemOpen [0]{}%
\providecommand \bibitemStop [0]{}%
\providecommand \bibitemNoStop [0]{.\EOS\space}%
\providecommand \EOS [0]{\spacefactor3000\relax}%
\providecommand \BibitemShut  [1]{\csname bibitem#1\endcsname}%
\let\auto@bib@innerbib\@empty
\bibitem [{\citenamefont {Loss}\ and\ \citenamefont
  {DiVincenzo}(1998)}]{Loss:1998p120}%
  \BibitemOpen
  \bibfield  {author} {\bibinfo {author} {\bibfnamefont {D.}~\bibnamefont
  {Loss}}\ and\ \bibinfo {author} {\bibfnamefont {D.~P.}\ \bibnamefont
  {DiVincenzo}},\ }\href@noop {} {\bibfield  {journal} {\bibinfo  {journal}
  {Phys Rev A}\ }\textbf {\bibinfo {volume} {57}},\ \bibinfo {pages} {120}
  (\bibinfo {year} {1998})}\BibitemShut {NoStop}%
\bibitem [{\citenamefont {Kane}(1998)}]{Kane:1998p133}%
  \BibitemOpen
  \bibfield  {author} {\bibinfo {author} {\bibfnamefont {B.~E.}\ \bibnamefont
  {Kane}},\ }\href@noop {} {\bibfield  {journal} {\bibinfo  {journal} {Nature}\
  }\textbf {\bibinfo {volume} {393}},\ \bibinfo {pages} {133} (\bibinfo {year}
  {1998})}\BibitemShut {NoStop}%
\bibitem [{\citenamefont {Elzerman}\ \emph
  {et~al.}(2004{\natexlab{a}})\citenamefont {Elzerman}, \citenamefont {Hanson},
  \citenamefont {van Beveren}, \citenamefont {Witkamp}, \citenamefont
  {Vandersypen},\ and\ \citenamefont {Kouwenhoven}}]{Elzerman:2004p431}%
  \BibitemOpen
  \bibfield  {author} {\bibinfo {author} {\bibfnamefont {J.~M.}\ \bibnamefont
  {Elzerman}}, \bibinfo {author} {\bibfnamefont {R.}~\bibnamefont {Hanson}},
  \bibinfo {author} {\bibfnamefont {L.~H.~W.}\ \bibnamefont {van Beveren}},
  \bibinfo {author} {\bibfnamefont {B.}~\bibnamefont {Witkamp}}, \bibinfo
  {author} {\bibfnamefont {L.~M.~K.}\ \bibnamefont {Vandersypen}}, \ and\
  \bibinfo {author} {\bibfnamefont {L.~P.}\ \bibnamefont {Kouwenhoven}},\
  }\href {\doibase 10.1038/nature02693} {\bibfield  {journal} {\bibinfo
  {journal} {Nature}\ }\textbf {\bibinfo {volume} {430}},\ \bibinfo {pages}
  {431} (\bibinfo {year} {2004}{\natexlab{a}})}\BibitemShut {NoStop}%
\bibitem [{\citenamefont {Petta}\ \emph {et~al.}(2005)\citenamefont {Petta},
  \citenamefont {Johnson}, \citenamefont {Taylor}, \citenamefont {Laird},
  \citenamefont {Yacoby}, \citenamefont {Lukin}, \citenamefont {Marcus},
  \citenamefont {Hanson},\ and\ \citenamefont {Gossard}}]{Petta:2005p2180}%
  \BibitemOpen
  \bibfield  {author} {\bibinfo {author} {\bibfnamefont {J.~R.}\ \bibnamefont
  {Petta}}, \bibinfo {author} {\bibfnamefont {A.~C.}\ \bibnamefont {Johnson}},
  \bibinfo {author} {\bibfnamefont {J.~M.}\ \bibnamefont {Taylor}}, \bibinfo
  {author} {\bibfnamefont {E.~A.}\ \bibnamefont {Laird}}, \bibinfo {author}
  {\bibfnamefont {A.}~\bibnamefont {Yacoby}}, \bibinfo {author} {\bibfnamefont
  {M.~D.}\ \bibnamefont {Lukin}}, \bibinfo {author} {\bibfnamefont {C.~M.}\
  \bibnamefont {Marcus}}, \bibinfo {author} {\bibfnamefont {M.~P.}\
  \bibnamefont {Hanson}}, \ and\ \bibinfo {author} {\bibfnamefont {A.~C.}\
  \bibnamefont {Gossard}},\ }\href {\doibase 10.1126/science.1116955}
  {\bibfield  {journal} {\bibinfo  {journal} {Science}\ }\textbf {\bibinfo
  {volume} {309}},\ \bibinfo {pages} {2180} (\bibinfo {year}
  {2005})}\BibitemShut {NoStop}%
\bibitem [{\citenamefont {Pioro-Ladri\`{e}re}\ \emph
  {et~al.}(2008)\citenamefont {Pioro-Ladri\`{e}re}, \citenamefont {Obata},
  \citenamefont {Tokura}, \citenamefont {Shin}, \citenamefont {Kubo},
  \citenamefont {Yoshida}, \citenamefont {Taniyama},\ and\ \citenamefont
  {Tarucha}}]{PioroLadriere:2008p776}%
  \BibitemOpen
  \bibfield  {author} {\bibinfo {author} {\bibfnamefont {M.}~\bibnamefont
  {Pioro-Ladri\`{e}re}}, \bibinfo {author} {\bibfnamefont {T.}~\bibnamefont
  {Obata}}, \bibinfo {author} {\bibfnamefont {Y.}~\bibnamefont {Tokura}},
  \bibinfo {author} {\bibfnamefont {Y.-S.}\ \bibnamefont {Shin}}, \bibinfo
  {author} {\bibfnamefont {T.}~\bibnamefont {Kubo}}, \bibinfo {author}
  {\bibfnamefont {K.}~\bibnamefont {Yoshida}}, \bibinfo {author} {\bibfnamefont
  {T.}~\bibnamefont {Taniyama}}, \ and\ \bibinfo {author} {\bibfnamefont
  {S.}~\bibnamefont {Tarucha}},\ }\href {\doibase doi:10.1038/nphys1053}
  {\bibfield  {journal} {\bibinfo  {journal} {Nat. Phys.}\ }\textbf {\bibinfo
  {volume} {4}},\ \bibinfo {pages} {776} (\bibinfo {year} {2008})}\BibitemShut
  {NoStop}%
\bibitem [{\citenamefont {MacLean}\ \emph {et~al.}(2007)\citenamefont
  {MacLean}, \citenamefont {Amasha}, \citenamefont {Radu}, \citenamefont
  {Zumb{\"u}hl}, \citenamefont {Kastner}, \citenamefont {Hanson},\ and\
  \citenamefont {Gossard}}]{MacLean:2007p1499}%
  \BibitemOpen
  \bibfield  {author} {\bibinfo {author} {\bibfnamefont {K.}~\bibnamefont
  {MacLean}}, \bibinfo {author} {\bibfnamefont {S.}~\bibnamefont {Amasha}},
  \bibinfo {author} {\bibfnamefont {I.~P.}\ \bibnamefont {Radu}}, \bibinfo
  {author} {\bibfnamefont {D.~M.}\ \bibnamefont {Zumb{\"u}hl}}, \bibinfo
  {author} {\bibfnamefont {M.~A.}\ \bibnamefont {Kastner}}, \bibinfo {author}
  {\bibfnamefont {M.~P.}\ \bibnamefont {Hanson}}, \ and\ \bibinfo {author}
  {\bibfnamefont {A.~C.}\ \bibnamefont {Gossard}},\ }\href {\doibase
  10.1103/PhysRevLett.98.036802} {\bibfield  {journal} {\bibinfo  {journal}
  {Phys. Rev. Lett.}\ }\textbf {\bibinfo {volume} {98}},\ \bibinfo {pages}
  {036802} (\bibinfo {year} {2007})}\BibitemShut {NoStop}%
\bibitem [{\citenamefont {Elzerman}\ \emph
  {et~al.}(2004{\natexlab{b}})\citenamefont {Elzerman}, \citenamefont {Hanson},
  \citenamefont {van Beveren}, \citenamefont {Vandersypen},\ and\ \citenamefont
  {Kouwenhoven}}]{Elzerman:2004p731}%
  \BibitemOpen
  \bibfield  {author} {\bibinfo {author} {\bibfnamefont {J.~M.}\ \bibnamefont
  {Elzerman}}, \bibinfo {author} {\bibfnamefont {R.}~\bibnamefont {Hanson}},
  \bibinfo {author} {\bibfnamefont {L.~H.~W.}\ \bibnamefont {van Beveren}},
  \bibinfo {author} {\bibfnamefont {L.~M.~K.}\ \bibnamefont {Vandersypen}}, \
  and\ \bibinfo {author} {\bibfnamefont {L.~P.}\ \bibnamefont {Kouwenhoven}},\
  }\href {\doibase 10.1063/1.1757023} {\bibfield  {journal} {\bibinfo
  {journal} {Appl. Phys. Lett.}\ }\textbf {\bibinfo {volume} {84}},\ \bibinfo
  {pages} {4617} (\bibinfo {year} {2004}{\natexlab{b}})}\BibitemShut {NoStop}%
\bibitem [{\citenamefont {Schleser}\ \emph {et~al.}(2005)\citenamefont
  {Schleser}, \citenamefont {Ruh}, \citenamefont {Ihn}, \citenamefont
  {Ensslin}, \citenamefont {Driscoll},\ and\ \citenamefont
  {Gossard}}]{Schleser:2005p035312}%
  \BibitemOpen
  \bibfield  {author} {\bibinfo {author} {\bibfnamefont {R.}~\bibnamefont
  {Schleser}}, \bibinfo {author} {\bibfnamefont {E.}~\bibnamefont {Ruh}},
  \bibinfo {author} {\bibfnamefont {T.}~\bibnamefont {Ihn}}, \bibinfo {author}
  {\bibfnamefont {K.}~\bibnamefont {Ensslin}}, \bibinfo {author} {\bibfnamefont
  {D.~C.}\ \bibnamefont {Driscoll}}, \ and\ \bibinfo {author} {\bibfnamefont
  {A.~C.}\ \bibnamefont {Gossard}},\ }\href@noop {} {\bibfield  {journal}
  {\bibinfo  {journal} {Phys. Rev. B}\ }\textbf {\bibinfo {volume} {72}},\
  \bibinfo {pages} {035312} (\bibinfo {year} {2005})}\BibitemShut {NoStop}%
\bibitem [{\citenamefont {Amasha}\ \emph
  {et~al.}(2008{\natexlab{a}})\citenamefont {Amasha}, \citenamefont {MacLean},
  \citenamefont {Radu}, \citenamefont {Zumb{\"u}hl}, \citenamefont {Kastner},
  \citenamefont {Hanson},\ and\ \citenamefont {Gossard}}]{Amasha:2008p1500}%
  \BibitemOpen
  \bibfield  {author} {\bibinfo {author} {\bibfnamefont {S.}~\bibnamefont
  {Amasha}}, \bibinfo {author} {\bibfnamefont {K.}~\bibnamefont {MacLean}},
  \bibinfo {author} {\bibfnamefont {I.~P.}\ \bibnamefont {Radu}}, \bibinfo
  {author} {\bibfnamefont {D.~M.}\ \bibnamefont {Zumb{\"u}hl}}, \bibinfo
  {author} {\bibfnamefont {M.~A.}\ \bibnamefont {Kastner}}, \bibinfo {author}
  {\bibfnamefont {M.~P.}\ \bibnamefont {Hanson}}, \ and\ \bibinfo {author}
  {\bibfnamefont {A.~C.}\ \bibnamefont {Gossard}},\ }\href {\doibase
  10.1103/PhysRevB.78.041306} {\bibfield  {journal} {\bibinfo  {journal} {Phys.
  Rev. B}\ }\textbf {\bibinfo {volume} {78}},\ \bibinfo {pages} {041306}
  (\bibinfo {year} {2008}{\natexlab{a}})}\BibitemShut {NoStop}%
\bibitem [{\citenamefont {Hanson}\ \emph {et~al.}(2005)\citenamefont {Hanson},
  \citenamefont {van Beveren}, \citenamefont {Vink}, \citenamefont {Elzerman},
  \citenamefont {Naber}, \citenamefont {Koppens}, \citenamefont {Kouwenhoven},\
  and\ \citenamefont {Vandersypen}}]{Hanson:2005p719}%
  \BibitemOpen
  \bibfield  {author} {\bibinfo {author} {\bibfnamefont {R.}~\bibnamefont
  {Hanson}}, \bibinfo {author} {\bibfnamefont {L.~H.~W.}\ \bibnamefont {van
  Beveren}}, \bibinfo {author} {\bibfnamefont {I.~T.}\ \bibnamefont {Vink}},
  \bibinfo {author} {\bibfnamefont {J.~M.}\ \bibnamefont {Elzerman}}, \bibinfo
  {author} {\bibfnamefont {W.~J.~M.}\ \bibnamefont {Naber}}, \bibinfo {author}
  {\bibfnamefont {F.~H.~L.}\ \bibnamefont {Koppens}}, \bibinfo {author}
  {\bibfnamefont {L.~P.}\ \bibnamefont {Kouwenhoven}}, \ and\ \bibinfo {author}
  {\bibfnamefont {L.~M.~K.}\ \bibnamefont {Vandersypen}},\ }\href {\doibase
  10.1103/PhysRevLett.94.196802} {\bibfield  {journal} {\bibinfo  {journal}
  {Phys. Rev. Lett.}\ }\textbf {\bibinfo {volume} {94}},\ \bibinfo {pages}
  {196802} (\bibinfo {year} {2005})}\BibitemShut {NoStop}%
\bibitem [{\citenamefont {Amasha}\ \emph
  {et~al.}(2008{\natexlab{b}})\citenamefont {Amasha}, \citenamefont {Maclean},
  \citenamefont {Radu}, \citenamefont {Zumb{\"u}hl}, \citenamefont {Kastner},
  \citenamefont {Hanson},\ and\ \citenamefont {Gossard}}]{Amasha:2008p1987}%
  \BibitemOpen
  \bibfield  {author} {\bibinfo {author} {\bibfnamefont {S.}~\bibnamefont
  {Amasha}}, \bibinfo {author} {\bibfnamefont {K.}~\bibnamefont {Maclean}},
  \bibinfo {author} {\bibfnamefont {I.}~\bibnamefont {Radu}}, \bibinfo {author}
  {\bibfnamefont {D.}~\bibnamefont {Zumb{\"u}hl}}, \bibinfo {author}
  {\bibfnamefont {M.}~\bibnamefont {Kastner}}, \bibinfo {author} {\bibfnamefont
  {M.}~\bibnamefont {Hanson}}, \ and\ \bibinfo {author} {\bibfnamefont
  {A.}~\bibnamefont {Gossard}},\ }\href {\doibase
  10.1103/PhysRevLett.100.046803} {\bibfield  {journal} {\bibinfo  {journal}
  {Phys. Rev. Lett.}\ }\textbf {\bibinfo {volume} {100}},\ \bibinfo {pages}
  {046803} (\bibinfo {year} {2008}{\natexlab{b}})}\BibitemShut {NoStop}%
\bibitem [{\citenamefont {Barthel}\ \emph {et~al.}(2009)\citenamefont
  {Barthel}, \citenamefont {Reilly}, \citenamefont {Marcus}, \citenamefont
  {Hanson},\ and\ \citenamefont {Gossard}}]{Barthel:2009p160503}%
  \BibitemOpen
  \bibfield  {author} {\bibinfo {author} {\bibfnamefont {C.}~\bibnamefont
  {Barthel}}, \bibinfo {author} {\bibfnamefont {D.~J.}\ \bibnamefont {Reilly}},
  \bibinfo {author} {\bibfnamefont {C.~M.}\ \bibnamefont {Marcus}}, \bibinfo
  {author} {\bibfnamefont {M.~P.}\ \bibnamefont {Hanson}}, \ and\ \bibinfo
  {author} {\bibfnamefont {A.~C.}\ \bibnamefont {Gossard}},\ }\href@noop {}
  {\bibfield  {journal} {\bibinfo  {journal} {Phys. Rev. Lett.}\ }\textbf
  {\bibinfo {volume} {103}},\ \bibinfo {pages} {160503} (\bibinfo {year}
  {2009})}\BibitemShut {NoStop}%
\bibitem [{\citenamefont {Xiao}\ \emph {et~al.}(2010)\citenamefont {Xiao},
  \citenamefont {House},\ and\ \citenamefont {Jiang}}]{Xiao:2010p1876}%
  \BibitemOpen
  \bibfield  {author} {\bibinfo {author} {\bibfnamefont {M.}~\bibnamefont
  {Xiao}}, \bibinfo {author} {\bibfnamefont {M.~G.}\ \bibnamefont {House}}, \
  and\ \bibinfo {author} {\bibfnamefont {H.~W.}\ \bibnamefont {Jiang}},\ }\href
  {\doibase 10.1103/PhysRevLett.104.096801} {\bibfield  {journal} {\bibinfo
  {journal} {Phys. Rev. Lett.}\ }\textbf {\bibinfo {volume} {104}},\ \bibinfo
  {pages} {096801} (\bibinfo {year} {2010})}\BibitemShut {NoStop}%
\bibitem [{\citenamefont {Hayes}\ \emph {et~al.}()\citenamefont {Hayes},
  \citenamefont {Kiselev}, \citenamefont {Borselli}, \citenamefont {Bui},
  \citenamefont {Croke}, \citenamefont {Deelman}, \citenamefont {Maune},
  \citenamefont {Milosavljevic}, \citenamefont {Moon}, \citenamefont {Ross},
  \citenamefont {Schmitz}, \citenamefont {Gyure},\ and\ \citenamefont
  {Hunter}}]{Hayes:2009preprint}%
  \BibitemOpen
  \bibfield  {author} {\bibinfo {author} {\bibfnamefont {R.~R.}\ \bibnamefont
  {Hayes}}, \bibinfo {author} {\bibfnamefont {A.~A.}\ \bibnamefont {Kiselev}},
  \bibinfo {author} {\bibfnamefont {M.~G.}\ \bibnamefont {Borselli}}, \bibinfo
  {author} {\bibfnamefont {S.~S.}\ \bibnamefont {Bui}}, \bibinfo {author}
  {\bibfnamefont {E.~T.}\ \bibnamefont {Croke}}, \bibinfo {author}
  {\bibfnamefont {P.~W.}\ \bibnamefont {Deelman}}, \bibinfo {author}
  {\bibfnamefont {B.~M.}\ \bibnamefont {Maune}}, \bibinfo {author}
  {\bibfnamefont {I.}~\bibnamefont {Milosavljevic}}, \bibinfo {author}
  {\bibfnamefont {J.-S.}\ \bibnamefont {Moon}}, \bibinfo {author}
  {\bibfnamefont {R.~S.}\ \bibnamefont {Ross}}, \bibinfo {author}
  {\bibfnamefont {A.~E.}\ \bibnamefont {Schmitz}}, \bibinfo {author}
  {\bibfnamefont {M.~F.}\ \bibnamefont {Gyure}}, \ and\ \bibinfo {author}
  {\bibfnamefont {A.~T.}\ \bibnamefont {Hunter}},\ }\href@noop {} {\enquote
  {\bibinfo {title} {Lifetime measurements ({T}$_1$) of electron spins in
  {S}i/{S}i{G}e quantum dots},}\ }\bibinfo {note} {ArXiv:0908.0173}\BibitemShut
  {NoStop}%
\bibitem [{\citenamefont {Huebl}\ \emph {et~al.}(2010)\citenamefont {Huebl},
  \citenamefont {Nugroho}, \citenamefont {Morello}, \citenamefont {Escott},
  \citenamefont {Eriksson}, \citenamefont {Yang}, \citenamefont {Jamieson},
  \citenamefont {Clark},\ and\ \citenamefont {Dzurak}}]{Huebl:2010p1868}%
  \BibitemOpen
  \bibfield  {author} {\bibinfo {author} {\bibfnamefont {H.}~\bibnamefont
  {Huebl}}, \bibinfo {author} {\bibfnamefont {C.~D.}\ \bibnamefont {Nugroho}},
  \bibinfo {author} {\bibfnamefont {A.}~\bibnamefont {Morello}}, \bibinfo
  {author} {\bibfnamefont {C.~C.}\ \bibnamefont {Escott}}, \bibinfo {author}
  {\bibfnamefont {M.~A.}\ \bibnamefont {Eriksson}}, \bibinfo {author}
  {\bibfnamefont {C.}~\bibnamefont {Yang}}, \bibinfo {author} {\bibfnamefont
  {D.~N.}\ \bibnamefont {Jamieson}}, \bibinfo {author} {\bibfnamefont {R.~G.}\
  \bibnamefont {Clark}}, \ and\ \bibinfo {author} {\bibfnamefont {A.~S.}\
  \bibnamefont {Dzurak}},\ }\href {\doibase 10.1103/PhysRevB.81.235318}
  {\bibfield  {journal} {\bibinfo  {journal} {Phys Rev B}\ }\textbf {\bibinfo
  {volume} {81}},\ \bibinfo {pages} {235318} (\bibinfo {year}
  {2010})}\BibitemShut {NoStop}%
\bibitem [{\citenamefont {Morello}\ \emph {et~al.}(2010)\citenamefont
  {Morello}, \citenamefont {Pla}, \citenamefont {Zwanenburg}, \citenamefont
  {Chan}, \citenamefont {Tan}, \citenamefont {Huebl}, \citenamefont {Mottonen},
  \citenamefont {Nugroho}, \citenamefont {Yang}, \citenamefont {van Donkelaar},
  \citenamefont {Alves}, \citenamefont {Jamieson}, \citenamefont {Escott},
  \citenamefont {Hollenberg}, \citenamefont {Clark},\ and\ \citenamefont
  {Dzurak}}]{Morello:2010p2645}%
  \BibitemOpen
  \bibfield  {author} {\bibinfo {author} {\bibfnamefont {A.}~\bibnamefont
  {Morello}}, \bibinfo {author} {\bibfnamefont {J.}~\bibnamefont {Pla}},
  \bibinfo {author} {\bibfnamefont {F.}~\bibnamefont {Zwanenburg}}, \bibinfo
  {author} {\bibfnamefont {K.}~\bibnamefont {Chan}}, \bibinfo {author}
  {\bibfnamefont {K.}~\bibnamefont {Tan}}, \bibinfo {author} {\bibfnamefont
  {H.}~\bibnamefont {Huebl}}, \bibinfo {author} {\bibfnamefont
  {M.}~\bibnamefont {Mottonen}}, \bibinfo {author} {\bibfnamefont
  {C.}~\bibnamefont {Nugroho}}, \bibinfo {author} {\bibfnamefont
  {C.}~\bibnamefont {Yang}}, \bibinfo {author} {\bibfnamefont {J.}~\bibnamefont
  {van Donkelaar}}, \bibinfo {author} {\bibfnamefont {A.}~\bibnamefont
  {Alves}}, \bibinfo {author} {\bibfnamefont {D.}~\bibnamefont {Jamieson}},
  \bibinfo {author} {\bibfnamefont {C.}~\bibnamefont {Escott}}, \bibinfo
  {author} {\bibfnamefont {L.}~\bibnamefont {Hollenberg}}, \bibinfo {author}
  {\bibfnamefont {R.}~\bibnamefont {Clark}}, \ and\ \bibinfo {author}
  {\bibfnamefont {A.}~\bibnamefont {Dzurak}},\ }\href@noop {} {\bibfield
  {journal} {\bibinfo  {journal} {Nature}\ }\textbf {\bibinfo {volume} {467}},\
  \bibinfo {pages} {687} (\bibinfo {year} {2010})}\BibitemShut {NoStop}%
\bibitem [{\citenamefont {Thalakulam}\ \emph {et~al.}(2010)\citenamefont
  {Thalakulam}, \citenamefont {Simmons}, \citenamefont {Rosemeyer},
  \citenamefont {Savage}, \citenamefont {Lagally}, \citenamefont {Friesen},
  \citenamefont {Coppersmith},\ and\ \citenamefont
  {Eriksson}}]{Thalakulam:2010p183104}%
  \BibitemOpen
  \bibfield  {author} {\bibinfo {author} {\bibfnamefont {M.}~\bibnamefont
  {Thalakulam}}, \bibinfo {author} {\bibfnamefont {C.~B.}\ \bibnamefont
  {Simmons}}, \bibinfo {author} {\bibfnamefont {B.~M.}\ \bibnamefont
  {Rosemeyer}}, \bibinfo {author} {\bibfnamefont {D.~E.}\ \bibnamefont
  {Savage}}, \bibinfo {author} {\bibfnamefont {M.~G.}\ \bibnamefont {Lagally}},
  \bibinfo {author} {\bibfnamefont {M.}~\bibnamefont {Friesen}}, \bibinfo
  {author} {\bibfnamefont {S.~N.}\ \bibnamefont {Coppersmith}}, \ and\ \bibinfo
  {author} {\bibfnamefont {M.~A.}\ \bibnamefont {Eriksson}},\ }\href@noop {}
  {\bibfield  {journal} {\bibinfo  {journal} {Appl. Phys. Lett.}\ }\textbf
  {\bibinfo {volume} {96}},\ \bibinfo {pages} {183104} (\bibinfo {year}
  {2010})}\BibitemShut {NoStop}%
\bibitem [{\citenamefont {Slinker}\ \emph {et~al.}(2005)\citenamefont
  {Slinker}, \citenamefont {Lewis}, \citenamefont {Haselby}, \citenamefont
  {Goswami}, \citenamefont {Klein}, \citenamefont {Chu}, \citenamefont
  {Coppersmith}, \citenamefont {Joynt}, \citenamefont {Blick}, \citenamefont
  {Friesen},\ and\ \citenamefont {Eriksson}}]{Slinker:2005p246}%
  \BibitemOpen
  \bibfield  {author} {\bibinfo {author} {\bibfnamefont {K.~A.}\ \bibnamefont
  {Slinker}}, \bibinfo {author} {\bibfnamefont {K.~L.~M.}\ \bibnamefont
  {Lewis}}, \bibinfo {author} {\bibfnamefont {C.~C.}\ \bibnamefont {Haselby}},
  \bibinfo {author} {\bibfnamefont {S.}~\bibnamefont {Goswami}}, \bibinfo
  {author} {\bibfnamefont {L.~J.}\ \bibnamefont {Klein}}, \bibinfo {author}
  {\bibfnamefont {J.~O.}\ \bibnamefont {Chu}}, \bibinfo {author} {\bibfnamefont
  {S.~N.}\ \bibnamefont {Coppersmith}}, \bibinfo {author} {\bibfnamefont
  {R.}~\bibnamefont {Joynt}}, \bibinfo {author} {\bibfnamefont {R.~H.}\
  \bibnamefont {Blick}}, \bibinfo {author} {\bibfnamefont {M.}~\bibnamefont
  {Friesen}}, \ and\ \bibinfo {author} {\bibfnamefont {M.~A.}\ \bibnamefont
  {Eriksson}},\ }\href {\doibase 10.1088/1367-2630/7/1/246} {\bibfield
  {journal} {\bibinfo  {journal} {New J. Phys.}\ }\textbf {\bibinfo {volume}
  {7}},\ \bibinfo {pages} {246} (\bibinfo {year} {2005})}\BibitemShut {NoStop}%
\bibitem [{\citenamefont {Berer}\ \emph {et~al.}(2006)\citenamefont {Berer},
  \citenamefont {Pachinger}, \citenamefont {Pillwein}, \citenamefont
  {M\"{u}hlberger}, \citenamefont {Lichtenberger}, \citenamefont {Brunthaler},\
  and\ \citenamefont {Sch\"{a}ffler}}]{Berer:2006p162112}%
  \BibitemOpen
  \bibfield  {author} {\bibinfo {author} {\bibfnamefont {T.}~\bibnamefont
  {Berer}}, \bibinfo {author} {\bibfnamefont {D.}~\bibnamefont {Pachinger}},
  \bibinfo {author} {\bibfnamefont {G.}~\bibnamefont {Pillwein}}, \bibinfo
  {author} {\bibfnamefont {M.}~\bibnamefont {M\"{u}hlberger}}, \bibinfo
  {author} {\bibfnamefont {H.}~\bibnamefont {Lichtenberger}}, \bibinfo {author}
  {\bibfnamefont {G.}~\bibnamefont {Brunthaler}}, \ and\ \bibinfo {author}
  {\bibfnamefont {F.}~\bibnamefont {Sch\"{a}ffler}},\ }\href {\doibase
  10.1063/1.2197320} {\bibfield  {journal} {\bibinfo  {journal} {Appl. Phys.
  Lett.}\ }\textbf {\bibinfo {volume} {88}},\ \bibinfo {pages} {162112}
  (\bibinfo {year} {2006})}\BibitemShut {NoStop}%
\bibitem [{\citenamefont {Ciorga}\ \emph {et~al.}(2000)\citenamefont {Ciorga},
  \citenamefont {Sachrajda}, \citenamefont {Hawrylak}, \citenamefont {Gould},
  \citenamefont {Zawadzki}, \citenamefont {Jullian}, \citenamefont {Feng},\
  and\ \citenamefont {Wasilewski}}]{Ciorga:2000p16315}%
  \BibitemOpen
  \bibfield  {author} {\bibinfo {author} {\bibfnamefont {M.}~\bibnamefont
  {Ciorga}}, \bibinfo {author} {\bibfnamefont {A.~S.}\ \bibnamefont
  {Sachrajda}}, \bibinfo {author} {\bibfnamefont {P.}~\bibnamefont {Hawrylak}},
  \bibinfo {author} {\bibfnamefont {C.}~\bibnamefont {Gould}}, \bibinfo
  {author} {\bibfnamefont {P.}~\bibnamefont {Zawadzki}}, \bibinfo {author}
  {\bibfnamefont {S.}~\bibnamefont {Jullian}}, \bibinfo {author} {\bibfnamefont
  {Y.}~\bibnamefont {Feng}}, \ and\ \bibinfo {author} {\bibfnamefont
  {Z.}~\bibnamefont {Wasilewski}},\ }\href {\doibase
  10.1103/PhysRevB.61.R16315} {\bibfield  {journal} {\bibinfo  {journal} {Phys.
  Rev. B}\ }\textbf {\bibinfo {volume} {61}},\ \bibinfo {pages} {R16315}
  (\bibinfo {year} {2000})}\BibitemShut {NoStop}%
\bibitem [{\citenamefont {Simmons}\ \emph {et~al.}(2007)\citenamefont
  {Simmons}, \citenamefont {Thalakulam}, \citenamefont {Shaji}, \citenamefont
  {Klein}, \citenamefont {Qin}, \citenamefont {Blick}, \citenamefont {Savage},
  \citenamefont {Lagally}, \citenamefont {Coppersmith},\ and\ \citenamefont
  {Eriksson}}]{Simmons:2007p213103}%
  \BibitemOpen
  \bibfield  {author} {\bibinfo {author} {\bibfnamefont {C.~B.}\ \bibnamefont
  {Simmons}}, \bibinfo {author} {\bibfnamefont {M.}~\bibnamefont {Thalakulam}},
  \bibinfo {author} {\bibfnamefont {N.}~\bibnamefont {Shaji}}, \bibinfo
  {author} {\bibfnamefont {L.~J.}\ \bibnamefont {Klein}}, \bibinfo {author}
  {\bibfnamefont {H.}~\bibnamefont {Qin}}, \bibinfo {author} {\bibfnamefont
  {R.~H.}\ \bibnamefont {Blick}}, \bibinfo {author} {\bibfnamefont {D.~E.}\
  \bibnamefont {Savage}}, \bibinfo {author} {\bibfnamefont {M.~G.}\
  \bibnamefont {Lagally}}, \bibinfo {author} {\bibfnamefont {S.~N.}\
  \bibnamefont {Coppersmith}}, \ and\ \bibinfo {author} {\bibfnamefont {M.~A.}\
  \bibnamefont {Eriksson}},\ }\href {\doibase 10.1063/1.2816331} {\bibfield
  {journal} {\bibinfo  {journal} {Appl. Phys. Lett.}\ }\textbf {\bibinfo
  {volume} {91}},\ \bibinfo {pages} {213103} (\bibinfo {year}
  {2007})}\BibitemShut {NoStop}%
\bibitem [{\citenamefont {Sakr}\ \emph {et~al.}(2005)\citenamefont {Sakr},
  \citenamefont {Jiang}, \citenamefont {Yablonovitch},\ and\ \citenamefont
  {Croke}}]{Sakr:2005p223104}%
  \BibitemOpen
  \bibfield  {author} {\bibinfo {author} {\bibfnamefont {M.~R.}\ \bibnamefont
  {Sakr}}, \bibinfo {author} {\bibfnamefont {H.~W.}\ \bibnamefont {Jiang}},
  \bibinfo {author} {\bibfnamefont {E.}~\bibnamefont {Yablonovitch}}, \ and\
  \bibinfo {author} {\bibfnamefont {E.~T.}\ \bibnamefont {Croke}},\ }\href
  {\doibase 10.1063/1.2136436} {\bibfield  {journal} {\bibinfo  {journal}
  {Appl. Phys. Lett.}\ }\textbf {\bibinfo {volume} {87}},\ \bibinfo {pages}
  {223104} (\bibinfo {year} {2005})}\BibitemShut {NoStop}%
\bibitem [{\citenamefont {Nordberg}\ \emph {et~al.}(2009)\citenamefont
  {Nordberg}, \citenamefont {Stalford}, \citenamefont {Young}, \citenamefont
  {Eyck}, \citenamefont {Eng}, \citenamefont {Tracy}, \citenamefont {Childs},
  \citenamefont {Wendt}, \citenamefont {Grubbs}, \citenamefont {Stevens},
  \citenamefont {Lilly}, \citenamefont {Eriksson},\ and\ \citenamefont
  {Carroll}}]{Nordberg:2009p202102}%
  \BibitemOpen
  \bibfield  {author} {\bibinfo {author} {\bibfnamefont {E.~P.}\ \bibnamefont
  {Nordberg}}, \bibinfo {author} {\bibfnamefont {H.~L.}\ \bibnamefont
  {Stalford}}, \bibinfo {author} {\bibfnamefont {R.}~\bibnamefont {Young}},
  \bibinfo {author} {\bibfnamefont {G.~A.~T.}\ \bibnamefont {Eyck}}, \bibinfo
  {author} {\bibfnamefont {K.}~\bibnamefont {Eng}}, \bibinfo {author}
  {\bibfnamefont {L.~A.}\ \bibnamefont {Tracy}}, \bibinfo {author}
  {\bibfnamefont {K.~D.}\ \bibnamefont {Childs}}, \bibinfo {author}
  {\bibfnamefont {J.~R.}\ \bibnamefont {Wendt}}, \bibinfo {author}
  {\bibfnamefont {R.~K.}\ \bibnamefont {Grubbs}}, \bibinfo {author}
  {\bibfnamefont {J.}~\bibnamefont {Stevens}}, \bibinfo {author} {\bibfnamefont
  {M.~P.}\ \bibnamefont {Lilly}}, \bibinfo {author} {\bibfnamefont {M.~A.}\
  \bibnamefont {Eriksson}}, \ and\ \bibinfo {author} {\bibfnamefont {M.~S.}\
  \bibnamefont {Carroll}},\ }\href@noop {} {\bibfield  {journal} {\bibinfo
  {journal} {Appl. Phys. Lett.}\ }\textbf {\bibinfo {volume} {95}},\ \bibinfo
  {pages} {202102} (\bibinfo {year} {2009})}\BibitemShut {NoStop}%
\bibitem [{\citenamefont {Simmons}\ \emph {et~al.}(2009)\citenamefont
  {Simmons}, \citenamefont {Thalakulam}, \citenamefont {Rosemeyer},
  \citenamefont {Bael}, \citenamefont {Sackmann}, \citenamefont {Savage},
  \citenamefont {Lagally}, \citenamefont {Joynt}, \citenamefont {Friesen},
  \citenamefont {Coppersmith},\ and\ \citenamefont
  {Eriksson}}]{Simmons:2009p3234}%
  \BibitemOpen
  \bibfield  {author} {\bibinfo {author} {\bibfnamefont {C.~B.}\ \bibnamefont
  {Simmons}}, \bibinfo {author} {\bibfnamefont {M.}~\bibnamefont {Thalakulam}},
  \bibinfo {author} {\bibfnamefont {B.~M.}\ \bibnamefont {Rosemeyer}}, \bibinfo
  {author} {\bibfnamefont {B.~J.~V.}\ \bibnamefont {Bael}}, \bibinfo {author}
  {\bibfnamefont {E.~K.}\ \bibnamefont {Sackmann}}, \bibinfo {author}
  {\bibfnamefont {D.~E.}\ \bibnamefont {Savage}}, \bibinfo {author}
  {\bibfnamefont {M.~G.}\ \bibnamefont {Lagally}}, \bibinfo {author}
  {\bibfnamefont {R.}~\bibnamefont {Joynt}}, \bibinfo {author} {\bibfnamefont
  {M.}~\bibnamefont {Friesen}}, \bibinfo {author} {\bibfnamefont {S.~N.}\
  \bibnamefont {Coppersmith}}, \ and\ \bibinfo {author} {\bibfnamefont {M.~A.}\
  \bibnamefont {Eriksson}},\ }\href@noop {} {\bibfield  {journal} {\bibinfo
  {journal} {Nano Lett.}\ }\textbf {\bibinfo {volume} {9}},\ \bibinfo {pages}
  {3234} (\bibinfo {year} {2009})}\BibitemShut {NoStop}%
\bibitem{Simmons:2010p245312}
  C. B. Simmons, T. S. Koh, N. Shaji, M. Thalakulam, L. J. Klein, H. Qin, H. Luo, D. E. Savage, M. G. Lagally, 
  A. J. Rimberg, R. Joynt, R. Blick, M. Friesen, S. N. Coppersmith, and M. A. Eriksson, \prb \textbf{82}, 
  245312 (2010).
\bibitem{Ono:2002p1313}
  K. Ono, D. G. Austing, Y. Tokura, and S. Tarucha, Science \textbf{297}, 1313 (2002).
\bibitem{Johnson:2005p925}
  A. C. Johnson, J. R. Petta, J. M. Taylor, A. Yacoby, M. D. Lukin, C. M. Marcus, M. P. Hanson, 
  and A. C. Gossard, Nature (London) \textbf{435}, 925 (2005)
\bibitem{Petta:2005p161301}
  J. R. Petta, A. C. Johnson, A. Yacoby, C. M. Marcus, M. P. Hanson, and A. C. Gossard, \prb \textbf{72}, 
  161301 (2005).
\bibitem{Johnson:2005p165308}
  A. C. Johnson, J. R. Petta, C. M. Marcus, M. P. Hanson, and A. C. Gossard, \prb \textbf{72}, 165308 (2005).
\bibitem [{\citenamefont {Kouwenhoven}\ \emph {et~al.}(1997)\citenamefont
  {Kouwenhoven}, \citenamefont {Marcus}, \citenamefont {McEuen}, \citenamefont
  {Tarucha}, \citenamefont {Westervelt},\ and\ \citenamefont
  {Wingreen}}]{Kouwenhoven:1997p1384}%
  \BibitemOpen
  \bibfield  {author} {\bibinfo {author} {\bibfnamefont {L.}~\bibnamefont
  {Kouwenhoven}}, \bibinfo {author} {\bibfnamefont {C.~M.}\ \bibnamefont
  {Marcus}}, \bibinfo {author} {\bibfnamefont {P.}~\bibnamefont {McEuen}},
  \bibinfo {author} {\bibfnamefont {S.}~\bibnamefont {Tarucha}}, \bibinfo
  {author} {\bibfnamefont {R.~M.}\ \bibnamefont {Westervelt}}, \ and\ \bibinfo
  {author} {\bibfnamefont {N.~S.}\ \bibnamefont {Wingreen}},\ }\enquote
  {\bibinfo {title} {Mesoscopic electron transport},}\ \ (\bibinfo  {publisher}
  {Kluwer, Dordrecht},\ \bibinfo {year} {1997})\BibitemShut {NoStop}%
\bibitem [{\citenamefont {Field}\ \emph {et~al.}(1993)\citenamefont {Field},
  \citenamefont {Smith}, \citenamefont {Pepper}, \citenamefont {Ritchie},
  \citenamefont {Frost}, \citenamefont {Jones},\ and\ \citenamefont
  {Hasko}}]{Field:1993p1477}%
  \BibitemOpen
  \bibfield  {author} {\bibinfo {author} {\bibfnamefont {M.}~\bibnamefont
  {Field}}, \bibinfo {author} {\bibfnamefont {C.~G.}\ \bibnamefont {Smith}},
  \bibinfo {author} {\bibfnamefont {M.}~\bibnamefont {Pepper}}, \bibinfo
  {author} {\bibfnamefont {D.~A.}\ \bibnamefont {Ritchie}}, \bibinfo {author}
  {\bibfnamefont {J.~E.~F.}\ \bibnamefont {Frost}}, \bibinfo {author}
  {\bibfnamefont {G.~A.~C.}\ \bibnamefont {Jones}}, \ and\ \bibinfo {author}
  {\bibfnamefont {D.~G.}\ \bibnamefont {Hasko}},\ }\href {\doibase
  10.1103/PhysRevLett.70.1311} {\bibfield  {journal} {\bibinfo  {journal}
  {Phys. Rev. Lett.}\ }\textbf {\bibinfo {volume} {70}},\ \bibinfo {pages}
  {1311} (\bibinfo {year} {1993})}\BibitemShut {NoStop}%
\bibitem [{\citenamefont {Hanson}\ \emph {et~al.}(2007)\citenamefont {Hanson},
  \citenamefont {Kouwenhoven}, \citenamefont {Petta}, \citenamefont {Tarucha},\
  and\ \citenamefont {Vandersypen}}]{Hanson:2007p720}%
  \BibitemOpen
  \bibfield  {author} {\bibinfo {author} {\bibfnamefont {R.}~\bibnamefont
  {Hanson}}, \bibinfo {author} {\bibfnamefont {L.~P.}\ \bibnamefont
  {Kouwenhoven}}, \bibinfo {author} {\bibfnamefont {J.~R.}\ \bibnamefont
  {Petta}}, \bibinfo {author} {\bibfnamefont {S.}~\bibnamefont {Tarucha}}, \
  and\ \bibinfo {author} {\bibfnamefont {L.~M.~K.}\ \bibnamefont
  {Vandersypen}},\ }\href {\doibase 10.1103/RevModPhys.79.1217} {\bibfield
  {journal} {\bibinfo  {journal} {Rev. Mod. Phys.}\ }\textbf {\bibinfo {volume}
  {79}},\ \bibinfo {pages} {1217} (\bibinfo {year} {2007})}\BibitemShut
  {NoStop}%
\bibitem{datta}  
  S. Datta, in \textit{Electronic Transport in Mesoscopic Systems} 
  (Cambridge University Press, Cambridge, 1995).
\end{thebibliography}
\end{document}